\documentclass{pasa}%
\usepackage{color}
\usepackage{amsmath}
\usepackage{amssymb}
\usepackage{natbib}
\usepackage{lscape}
\usepackage{gensymb}
\usepackage{verbatim} 
\usepackage{tablefootnote}
\DeclareGraphicsExtensions{.eps,.ps,.pdf,.jpg,.jpeg}
\usepackage{lastpage}
\usepackage{graphicx}
\usepackage{epsfig}
\newcommand{\msun}{M$_\odot$}
\newcommand{\arcmin}{$'$}
\newcommand{\arcsec}{$''$}

\def\Victoria{$^{1}$}
\def\Herts{$^{2}$}
\def\ICRAR{$^{3}$}
\def\USydney{$^{4}$}
\def\CAASTRO{$^{5}$}
\def\CASS{$^{6}$}
\def\SKASA{$^{7}$}
\def\CfA{$^{8}$}
\def\RU{$^{9}$}
\def\ASU{$^{10}$}
\def\ANU{$^{11}$}
\def\Haystack{$^{12}$}
\def\RRI{$^{13}$}
\def\ICRAUW{$^{14}$}
\def\UToronto{$^{15}$}
\def\UW{$^{16}$}
\def\UWisc{$^{17}$}
\def\UMelbourne{$^{18}$}
\def\MIT{$^{19}$}
\def\Tata{$^{20}$}
\def\Astron{$^{21}$}
\title[Low Frequency HII regions]{A Large Scale, Low Frequency Murchison Widefield Array Survey of Galactic HII regions between 260$\,<\,$\textit{l}$\,<\,$340}
\author[Hindson et al.]{L.~Hindson\Victoria$^,$\Herts, M.~Johnston-Hollitt\Victoria, N.~ Hurley-Walker\ICRAR, J.~R.~Callingham\USydney$^,$\CAASTRO$^,$\CASS, H.~Su\ICRAR, J.~Morgan\ICRAR, M.~Bell\CAASTRO$^,$\CASS, G.~Bernardi\SKASA$^,$\CfA$^,$\RU, J.~D.~Bowman\ASU, F.~Briggs\ANU, R.~J.~Cappallo\Haystack, A.~A.~Deshpande\RRI, K.~S.~Dwarakanath\RRI, B.-Q~For\ICRAUW, B.~M.~Gaensler\UToronto$^,$\USydney$^,$\CAASTRO, L.~J.~Greenhill\CfA, P.~Hancock\ICRAR$^,$\CAASTRO, B.~J.~Hazelton\UW, A.~D.~Kapi\'{n}ska\ICRAUW$^,$\CAASTRO, D.~L.~Kaplan\UWisc, E.~Lenc\USydney$^,$\CAASTRO, C.~J.~Lonsdale\Haystack, B.~Mckinley\CAASTRO$^,$\ANU, S.~R.~McWhirter\Haystack, D.~A.~Mitchell\UMelbourne$^,$\CAASTRO$^,$\CASS, M.~F.~Morales\UW, E.~Morgan\MIT, D.~Oberoi\Tata, A.~Offringa\Astron, S.~M.~Ord\ICRAR$^,$\CAASTRO, P.~Procopio\UMelbourne$^,$\CAASTRO, T.~Prabu\RRI, N.~Udaya~Shankar\RRI, K.~S.~Srivani\RRI, L.~Staveley-Smith\ICRAUW$^,$\CAASTRO,  R.~Subrahmanyan\RRI$^,$\CAASTRO, S.~J.~Tingay\ICRAR$^,$\CAASTRO, R.~B.~Wayth\ICRAR$^,$\CAASTRO, R.~L.~Webster\UMelbourne$^,$\CAASTRO, A.~Williams\ICRAR, C.~L.~Williams\MIT, C.~Wu\ICRAUW, Q.~Zheng\Victoria\\
\\
\affil{$^{1}$School of Chemical \& Physical Sciences, Victoria University of Wellington, Wellington 6140, New Zealand}%
\affil{$^{2}$Centre for Astrophysics Research, School of Physics, Astronomy and
Mathematics, University of Hertfordshire, College Lane, Hatfield AL10
9AB, UK}%
\affil{$^{3}$International Centre for Radio Astronomy Research, Curtin University, Bentley, WA 6102, Australia}%
\affil{$^{4}$Sydney Institute for Astronomy, School of Physics, The University of Sydney, NSW 2006, Australia}%
\affil{$^{5}$ARC Centre of Excellence for All-sky Astrophysics (CAASTRO), Redfern, NSW, Australia}%
\affil{$^{6}$CSIRO Astronomy and Space Science (CASS), PO Box 76, Epping, NSW 1710, Australia}%
\affil{$^{7}$Square Kilometre Array South Africa (SKA SA), Cape Town 7405, South Africa }
\affil{$^{8}$Harvard-Smithsonian Center for Astrophysics, Cambridge, MA 02138, USA}%
\affil{$^{9}$Department of Physics and Electronics, Rhodes University, PO Box 94, Grahamstown 6140, South Africa}%
\affil{$^{10}$School of Earth and Space Exploration, Arizona State University, Tempe, AZ 85287, USA}%
\affil{$^{11}$Research School of Astronomy and Astrophysics, Australian National University, Canberra, ACT 2611, Australia}%
\affil{$^{12}$MIT Haystack Observatory, Westford, MA 01886, USA}%
\affil{$^{13}$Raman Research Institute, Bangalore 560080, India}%
\affil{$^{14}$International Centre for Radio Astronomy Research, University of Western Australia, Crawley, WA 6009, Australia}%
\affil{$^{15}$Dunlap Institute for Astronomy and Astrophysics, University of Toronto, ON, M5S 3H4, Canada}
\affil{$^{16}$Department of Physics, University of Washington, Seattle, WA 98195, USA}%
\affil{$^{17}$Department of Physics, University of Wisconsin--Milwaukee, Milwaukee, WI 53201, USA}%
\affil{$^{18}$School of Physics, The University of Melbourne, Parkville, VIC 3010, Australia}%
\affil{$^{19}$Kavli Institute for Astrophysics and Space Research, Massachusetts Institute of Technology, Cambridge, MA 02139, USA}%
\affil{$^{20}$National Centre for Radio Astrophysics, Tata Institute for Fundamental Research, Pune 411007, India}%
\affil{$^{21}$Netherlands Institute for Radio Astronomy (ASTRON), PO Box 2, 7990 AA Dwingeloo, The Netherlands }
}%

\jid{PASA}
\doi{10.1017/pas.\the\year.xxx}
\jyear{\the\year}

\usepackage{natbib}
\bibpunct{(}{)}{;}{a}{}{,}
\begin{document}%
\begin{abstract}
We have compiled a catalogue of HII regions detected with the Murchison Widefield Array (MWA) between 72 and 231\,MHz. The multiple frequency bands provided by the MWA allow us identify the characteristic spectrum generated by the thermal Bremsstrahlung process in HII regions. We detect 302 HII regions between $260$\degree$ < l < 340$\degree and report on the positions, sizes, peak, integrated flux density, and spectral indices of these HII regions. By identifying the point at which HII regions transition from the optically thin to thick regime we derive the physical properties including the electron density, ionised gas mass and ionising photon flux, towards 61 HII regions. This catalogue of HII regions represents the most extensive and uniform low frequency survey of HII regions in the Galaxy to date.
\end{abstract}
\maketitle%
\section{Introduction }
\label{sec:intro}

Observations of the Galactic plane in the radio regime are dominated by five distinct components: HII regions; supernova remnants (SNRs); the diffuse Galactic synchrotron background; pulsar wind nebulae, and background radio galaxies. Stars of spectral type $\sim$B0 or earlier (masses $>8$\msun) produce a sufficient number of high-energy ultra-violet photons to ionise a large region of the surrounding interstellar medium (ISM). These regions of ionised gas are known as HII regions and they can have a significant impact on the surrounding ISM through heating, photo-evaporation and dissipation, and the formation of D-type shocks \citep{Yorke1986}. HII regions are therefore an important driver of the chemical and kinematic evolution in galaxies and may be responsible for triggering new episodes of star formation \citep{Elmegreen1977,Thompson2012}. The massive stars responsible for generating HII regions have relatively short lifetimes on the order of $\sim 10$\,Myr and so provide an unambiguous tracer of the current epoch of massive star formation \citep{Zinnecker2007}. At the end of their lives massive stars undergo core collapse resulting in a supernova and associated remnant, which emits in the radio regime. The diffuse Galactic synchrotron background is generated by the emission of cosmic ray electrons (CRes) interacting with magnetic fields. It is now widely accepted that these CRes are initially accelerated in supernova events and then proceed to propagate outwards into the ISM through diffusion and convection \citep{Reynolds2008,Duric1999}. In addition to these Galactic sources of radio continuum emission, extragalactic sources comprising mainly radio galaxies are also observed. 

One distinct characteristic of HII regions compared to other sources of radio emission is their spectral index ($\alpha$: $S_{\rm \nu} \propto \nu^{\alpha}$). The radio continuum emission of HII regions is generated by the thermal Bremsstrahlung mechanism \citep{Rybicki1986}. The radio spectral index of this process depends on the properties of the HII region. For example in compact and classical HII regions with sizes from $s\approx0.5$ to 10\,pc; electron temperatures of $T_{\rm e}\approx10^4$\,K; and electron densities of $n_{\rm e}\approx 10^7$ to $10^2$cm$^{-3}$ the radio emission is optically thin above $\sim 200$\,MHz with a spectral index of $\alpha \approx -0.1$. Below this frequency such HII regions become optically thick leading to a turnover and rapid steeping of the spectral index with values of $\sim 2.0$ \citep{Kurtz2005,Mezger1967}. In contrast, the radio emission associated with the diffuse Galactic background, SNRs, and extragalactic sources is generated by the synchrotron process. This process leads to a much steeper negative spectral index in the optically thin regime than thermal Bremsstrahlung. The average spectral index of the diffuse background synchrotron emission in the Galactic plane is $\alpha\approx-0.8$ \citep{Platania1998} with typical values for SNRs of $\alpha\approx-0.5$ \citep{Green2014}, background radio galaxies of $\alpha \approx -0.85$ \citep{Lisenfeld2000}, and pulsar wind nebulae of $\alpha \approx 0.0$ to $-0.3$ \citep{Weiler1988}. There are three cases where the spectral index of SNRs may result in a flatter or positive spectral index. Firstly, all SNRs will turnover into the optically thick regime due to synchrotron self-absorption at very low  frequency ($\sim 1$\,MHz: \citealt{Ginzburg1969}). At frequencies below $\sim 100$\,MHz the SED of some SNRs have been found to turnover into the optically thick regime due to the presence of foreground absorbing gas \citep{Kassim1995}. Finally, the spectral index of very young and resolved SNRs have been found to flatten towards the central regions due to absorption by thermal ejecta associated with the SNR \citep{Delaney2014}. If consideration is given to these cases the spectral index can be used as a tool for identifying HII regions in the Galactic plane.

Surveys for HII regions have tended to be carried out using tracers such as their mid-infrared (MIR) colour and morphology, which probes the emission of heated dust \citep{Anderson2014,Churchwell2006,Urquhart2008}. Observations of hydrogen radio recombination lines are also commonly used \citep{Caswell1987,Bania2010,Bania2012}. Wide area radio continuum surveys of HII regions have been primarily carried out at frequencies above 1\,GHz and concentrated on searching for young and compact HII regions. These studies include: the Coordinated Radio and Infrared Survey for High-Mass Star Formation (CORNISH) survey at 5\,GHz \citep{Hoare2012}, the combined 20\,GHz and 843\,MHz study by \cite{Murphy2010}, and the 1.4 and 5\,GHz study carried out by \cite{Giveon2005}. Surveys such as these can be effectively used to identify HII regions but require the use of ancillary data, primarily infra-red observations, to positively identify HII regions. The low frequency regime and multiple frequency bands of the Murchison Widefield Array (MWA) provide an opportunity to uniformly survey large areas of the Galactic plane and identify optically thick HII regions by their spectral index and characterise their SEDs at low frequency.

Studies of the Galactic plane at low frequency are limited and have historically had poor angular resolution resulting in HII regions being poorly characterised below 1\,GHz. One of the most sensitive radio surveys below 1\,GHz is the Molonglo Galactic Plane Survey (MGPS) I and II at 843\,MHz \citep{Green1999,Murphy2007}. The MGPS catalogue has good coverage ($245$\degree$ < l < 365$\degree), resolution ($\sim 45$\arcsec), and sensitivity (1--2\,mJy\,beam$^{-1}$). Cross matching between the MGPS survey and MIR data from the Midcourse Space Experiment (MSX: \citealt{Mill1994}) has been used by \cite{Cohen2001} to identify HII regions. However, MGPS images suffer from significant artefacts, a limited bandwidth (3\,MHz), and poor spatial sensitivity that limits the ability to sample angular scales above $\sim 30$\arcmin. This makes characterising HII regions difficult and prone to underestimating the total flux density. The Southern Galactic Plane Survey (SGPS: \citealt{Haverkorn2006}) observed the Galactic plane between $253$\degree$ < l < 358$\degree\ at 1420\,MHz with a resolution of 100\arcsec\ and sensitivity of 0.3--0.6\,mJy\,beam$^{-1}$. Whilst this is not typically characterised as the low frequency regime we mention the survey here as we make use of it later in Section~\ref{subsect:thick}. Like the MGPS, the spatial sensitivity of the SGPS is limited to structures $<30$\arcmin. The inner Galaxy has been studied at 74\,MHz using the Very Large Array (VLA) in multiple configurations \citep{Nord2006} (hereafter VLA\,74). These observations have good resolution and sensitivity ($\sim 10$\arcmin) but only mapped a small region of between 26\degree$> l > -15$\degree. The VLA has also surveyed the northern sky at 74\,MHz in the VLA Low-frequency Sky Survey (VLSS), which was re-reduced as the VLA Sky Survey Redux (VLSSr; \citealt{Lane2014}). The VLSSr has good sensitivity and resolution (0.1\,Jy\,beam$^{-1}$ and 75\arcsec) but lacks sensitivity to sources on spatial scales above $\sim23$\arcmin. The 408\,MHz survey of \cite{Haslam1982} mapped the entire Galactic plane using the single-dish Parkes and Jodrell Bank MK1A telescopes. This means that all spatial scales are sampled, however studies of individual HII regions are not possible due to the large $51$\arcmin\ beam. The Westerbork Galactic Plane Survey at 327\,MHz (WENSS: \citealt{Taylor1996}) and the Canadian Galactic Plane Survey (CGPS: \citealt{Taylor2003}) at 408\,MHz  have carried out surveys of the northern Galactic plane with a resolution of $\sim1$ and $3$\arcmin\ and sensitivities of 10 and 3\,mJy\,beam$^{-1}$, respectively. The CGPS is an interferometric study that made use of the single-dish maps of \cite{Haslam1982} to recover spatial information on large scales. There has not yet been a low frequency survey of the southern Galactic plane of sufficient resolution, bandwidth, and sensitivity to large scale structure to study HII regions in detail at low frequency. In particular, the limited bandwidth and spatial sensitivity of surveys such as MGPS and SGPS are unable to study the SED of Galactic emission. 

\begin{figure*}
\centering
\includegraphics[width=1.0\textwidth]{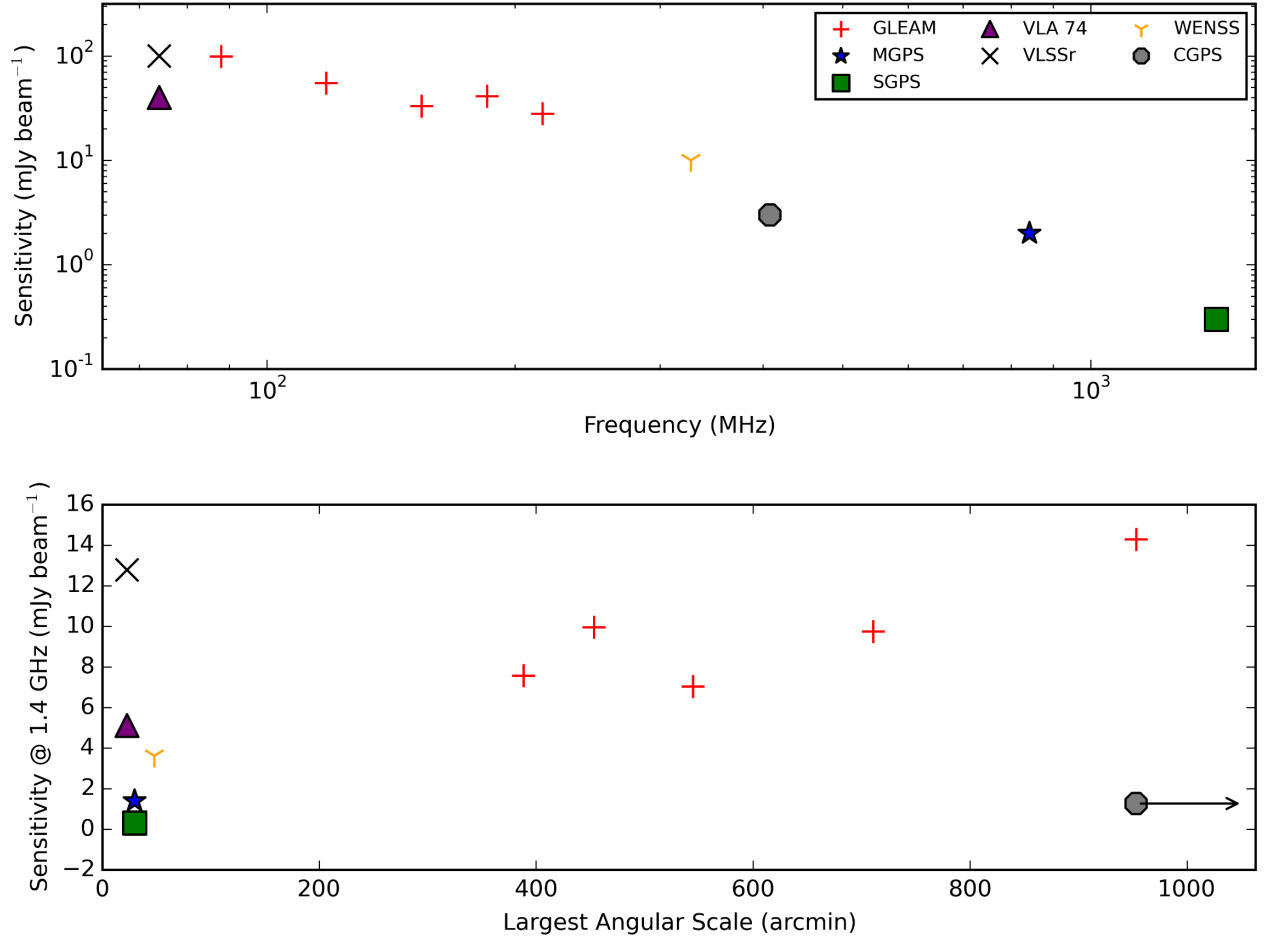} 
\caption{Top panel: this plot shows the logarithm of the frequency vs. sensitivity for low frequency Galactic surveys ($<1.4$\,GHz). Bottom panel: the spatial sensitivity and corresponding sensitivity, scaled to 1.4\,GHz assuming a spectral index of $-0.7$. The CGPS survey includes single dish data and so recovers all spatial information. The MWA provides both high sensitivity and access to a broader range of angular scales compared with previous surveys of the Galactic plane.}
\label{im:Survey}
\end{figure*}

Characterising the low frequency SED of HII regions can provide important constraints on physical properties, such as the electron density and ionising flux, by identifying the turnover frequency, which is the point at which the HII region becomes optically thick \citep{Mezger1967}. These optically thick HII regions can also be seen as absorption features against the extended CRe synchrotron Galactic background. This allows the emissivity of CRes behind the HII region to be derived and can also be used to study the three-dimensional relativistic gas content in the Galaxy \citep{Sun2008,Nord2006}. In some cases numerous optically thick HII regions can lead to a drop in the synchrotron dominated integrated radio continuum of a galaxy, which may have important implications for models of starburst galaxies \citep{Lacki2013,Kassim1989}.

Recent developments in instrumentation are for the first time allowing wide-field, high resolution, and sensitive observations at low frequency. Instruments such as the GMRT \citep{Swarup1991}, JVLA \citep{Clarke2015}, LOFAR \citep{vanHaarlem2013}, and the MWA have either performed or are in the process of performing large-area sky surveys at low frequency. In this paper we present a technique for identifying HII regions using multi-band observations at low frequency. The resultant catalogue of HII regions was compiled using the GaLactic Extragalactic All-sky MWA (GLEAM: \citealt{Wayth2015}) survey between 72 and 231\,MHz. The GLEAM observations allow us to probe new areas of parameter space as compared with existing Galactic plane surveys (see Figure~\ref{im:Survey}). In particular using the MWA we are able to examine a largely unexplored frequency range on large spatial scales, making this an ideal work to detect HII regions. The paper is organised as follows: in Section~\ref{sec:obs} we discuss our observations and data analysis process. In Section~\ref{sec:finding} we present our method for detecting and characterising HII regions and provide a sample of our catalogue in Section~\ref{sec:results}. We present our discussion of these results in Section~\ref{sec:disc}, which includes: uncertainties, the physical properties of HII regions that show signs of spectral turnover; and a comparison to the HII region catalogue constructed by \cite{Anderson2014}. 

\section{Observations and Data Analysis}
\label{sec:obs}

We use observations that utilise the MWA interferometer, which is the low frequency precursor for the Square Kilometre Array (SKA), located at the Murchison Radio-astronomy Observatory in Western Australia. We provide a brief description of the MWA here and direct the reader to \cite{Tingay2013} and \cite{Bowman2013} for a detailed description of the MWA and its science goals. The MWA consists of 128 ``tiles'' of 16 dipoles, each with two polarisations, which are distributed in a dense core $<1.5$\,km in diameter with a sparser distribution of tiles out to a diameter of approximately 2.5\,km. This layout coupled with the very large number of antenna elements gives the array excellent snapshot $u$,$v$ coverage and sensitivity to low surface brightness emission. The field of view (FoV) of the MWA is 610 to 375\,deg$^2$ with a resolution from $\sim6$ to 2\arcmin\ and spatial sensitivity to structures from $\sim3$ to 950\arcmin, depending on frequency. The MWA operates at frequencies between 72 and 300\,MHz with an instantaneous bandwidth of 30.72\,MHz.

	\subsection{The GLEAM Survey}
The observations presented in this study were obtained as part of the GLEAM survey \citep{Wayth2015}. The GLEAM survey has observed the entire southern sky from a declination ($\delta$) of $+25$\degree\ to $-90$\degree. Observations were obtained using drift-scans at a range of declination settings ($\delta = +18.6^{\circ}$, $+1.6^{\circ}$, $-13.0^{\circ}$, $-26.7^{\circ}$, $-40.0^{\circ}$, $-55.0^{\circ}$, $-72.0^{\circ}$). Snapshots of 112\,s were collected in five 30.72\,MHz wide frequency bands between 72 and 231\,MHz centred on 87.67, 118.40, 154.24, 184.96, and 215.68\,MHz (hereafter 88, 118, 154, 185, and 216\,MHz). Scans of bright compact sources were obtained every two hours for use as calibrators. This results in 55 snapshots per frequency per night. The $\delta=-55.0$\degree\ observations used in this study were obtained on the 17$^{\rm th}$ of March 2014 and cover the sky from an RA of 6--16h. 

\begin{table}
\small
\caption[Observation characteristics]{Summary of the image properties for each of our MWA Galactic plane mosaics.}
\begin{center}
  \begin{tabular}{ccc}
   \hline Centre Frequency & Resolution & Sensitivity\\
           (MHz) & ($\arcmin$) & (mJy\,beam$^{-1}$) \\
 	\hline
88	&	$5.6\times5.1$	&	99	\\
118	&	$4.0\times3.7$	&	55	\\
154	&	$3.2\times2.9$	&	33	\\
185	&	$2.8\times2.5$	&	41	\\
216	&	$2.5\times2.3$	&	28	\\
\hline
\end{tabular}
\end{center}
\label{tab:obs}
\end{table}

	\subsection{Data Reduction}
The data reduction process used to flag, calibrate, and image individual snapshots follows a process very similar to that used in the MWA commissioning survey presented in \cite{Hurley2014}. The post processing steps, which includes establishing a flux density scale and mosaicking, follows the method utilised in the GLEAM survey pipeline \citep{Wayth2015,Hurley2015}. 

For each snapshot the raw data from the correlator were first processed through the {\sc cotter} pipeline \citep{Offringa2015}. This is a pre-processing pipeline designed to perform initial flagging and radio frequency interference excision using {\sc aoflagger}, data-averaging to 1\,s time and 40\,kHz frequency resolution, and conversion of the data into the measurement set format readable by {\sc casa\footnote{http://casa.nrao.edu/}}. A single complex gain solution was determined for each tile and frequency using the {\sc bandpass} routine on a single pointed observation of Hydra\,A. The calibrated Stokes XX and YY snapshots were then imaged using the {\sc clean} task with a robust weighting of 0.0. The pixel size for each image is set to 0.75\arcmin. This results in the synthesised beam being sampled by three pixels in the highest frequency image and leads to an oversampling of the synthesised beam at the lowest frequency by approximately seven pixels.

The poor sampling of low frequency radio sources complicates the characterisation of the flux density scale in our 55 snapshots. The only large scale survey carried out at these declinations is the Molonglo Reference Catalog (MRC) survey at 408\,MHz \citep{Large1981}. To define the flux density scale in our snapshots we first perform source finding using the {\sc aegean} algorithm \citep{Hancock2012} to produce a catalogue of sources in each Stokes XX and YY snapshot. We remove faint and extended sources and cross match the resulting catalogue of bright and unresolved sources with the MRC catalogue. This results in $\sim200$ sources per snapshot that we use to characterise the flux density scale. To determine the correction we need to apply to the flux density scale in each MWA snapshot, we first scale the MRC flux density to the appropriate MWA central frequency by assuming a spectral index of $-0.85$. The applied correction is the average of the $\sim 200$ sources weighted by the MWA source peak flux density divided by the local noise (rms). This provides an initial flux density scale in each snapshot that we refine in an additional step described below.

At low frequency the ionosphere generates slow astrometric changes in source position that can result in shifts on the order of 10--20\arcsec\ in 112\,s GLEAM snapshots at 154\,MHz \citep{Loi2015}. In an effort to correct for these shifts in the apparent position of sources we cross match the position of compact sources detected in each of the GLEAM snapshots with MRC sources to determine an average astrometric correction and update the headers in each snapshot accordingly. This method results in a residual offset in source position and is discussed further in Section~\ref{subsect:Uncertainties}.

The topology and wide FoV of the MWA means that the $w$-term in the standard two-dimensional visibility equation can no longer be assumed to be zero \citep{Thompson1999}. The varying $w$-term in each snapshot leads to a systematic positional offset increasing with distance from the centre of the observed field. To account for this effect the coordinate system of each sine projected snapshot was corrected by adding fits keywords to the header that allow us to represent the slant orthographic correction \citep{Perley1999,Calabretta2002}. The effect of $w$-projection also leads to sources being smeared out on a small scale. As sources appear further from the phase centre this effect becomes more apparent. Our reduction method is unable to account for this effect as at the time of the analysis the {\rm CASA} implementation of $w$-term correction was found to be far too computationally expensive. However, the Galactic plane is well centred in the primary beam in the majority of our snapshots and we primarily study resolved sources where the effects of $w$-term smearing are minimised. 

After applying these corrections we combine the stokes XX and YY snapshots in each band into large scale mosaics using {\sc swarp} \citep{Bertin2002}. Analysis of large scale GLEAM mosaics such as those produced here has revealed residual variation in the flux density with declination \citep{Hurley2015} . This is caused by errors of order 5--20\% in the primary beam model of the MWA \citep{Sutinjo2015}. To correct for this variation and refine the flux density scale, we adopted the strategy that is used in the GLEAM survey. This process involves identifying a sample of sources that are 8$\sigma$ above the noise floor of the MWA mosaic and brighter than 2\,Jy in VLSSr. These sources must also be: detected in MRC and NRAO/VLA Sky Survey (NVSS: \citealt{Condon1998}); unresolved at all the observed frequencies; isolated from potentially confusing sources including the Galactic plane; and have a spectrum that was well fit by a power law. Correction factors were derived by comparing the measured MWA flux density and the expected flux density derived from the power law fit to a source's spectrum. A polynomial was then fit to the correction factors as a function of declination to remove the residual primary beam uncertainties. Since the necessary frequency coverage of VLSSr is not available at the low declination of the $\delta=-55.0$\degree\ observations, we exploited the symmetrical nature of the MWA beam to derive the flux density dependant correction in the $\delta=+1.6$\degree\ drift scan and mirrored the results to the $\delta=-55.0$\degree\ region. Note that in this process all surveys used were placed on the absolute flux density scale of \cite{Baars1977}. Due to the slight variation in the correction factors, our flux density measurements have a systematic uncertainty of $\sim 5$--$8\%$.

Combining wide-field snapshots that have residual ionospheric and $w$-term errors results in a distortion of sources and a blurring of the point spread function
(PSF). This blurring of the PSF varies across the mosaic depending on the ionospheric conditions at the time and also the location of sources within the primary beam. The magnitude of these variations is 10--30\% and leads to a decrease in peak flux measurements in our mosaics. Again we correct for this following the same approach as the GLEAM survey. We characterise the PSF across our images by selecting unresolved and isolated MRC sources and fit the sources with a Gaussian model. We then use these results to map the position dependant PSF and correct the peak fluxes in our images accordingly. We summarise the resolution and sensitivity in our MWA mosaics in Table~\ref{tab:obs}. Given the angular resolution of these observations, $\sim 5.3$--$2.3$\arcmin, we expect to resolve HII regions with physical sizes $>1$--31\,pc at distances from 1--20\,kpc, respectively. The sensitivity to angular scales up to $\sim 950$--390\arcmin\ results in sensitivity to structures of $< 270$ and $< 5400$\,pc at distances of 1--20\,kpc. This allows us to recover flux from the largest of HII regions ($\sim 100$\,pc: \citealt{Kurtz2005}) at all bands which makes spectral index studies possible. With the exception of the CGPS in the northern sky, previous surveys have had to sacrifice good spatial sensitivity for resolution or have had limited Galactic coverage.

\section{Source-finding}
\label{sec:finding}

To identify and define the boundaries of features in our Galactic plane images we employ the {\sc fellwalker} clump finding algorithm \citep{Berry2015}, which is part of the {\sc starlink} project. {\sc fellwalker} is an automated thresholding approach to source detection that identifies contiguous features in an image by finding the paths of steepest gradient for each pixel. Starting with the first pixel in an image each of the surrounding pixels is inspected to locate the pixel with the highest ascending gradient. This process continues until a peak is located (i.e. a pixel surrounded by flat or descending gradients). The pixels along the steepest path to the peak are assigned an arbitrary integer to represent their connection along a path. All pixels in the image are inspected in a similar process and the image is segmented into clumps by grouping together all paths that lead to the same peak pixel. The output of this process is an image with the identified regions specified by their clump number which can be used as a template to perform aperture photometry.

We chose to use {\sc fellwalker} over other thresholding algorithms such as {\sc clumpfind} \citep{Williams1994} and {\sc blobcat} \citep{Hales2012} due to our familiarity and successful application of the algorithm to identify similarly complex emission in the Galactic plane \citep{Hindson2010}. In addition {\sc fellwalker} makes no assumptions as to the shape of a source, which makes it suited to characterising complex emission features. Source finding algorithms currently in use such as those above work in fundamentally the same way by applying a thresholding approach to source detection. There are very few, if any, automatic source detection algorithms that are able to robustly characterise complex or occluded sources (see \citealt{Hollitt2012} for a discussion). Regarding {\sc fellwalker} the gradient-based approach allows it to deal reasonably well with complex topologies and it allows easy manual manipulation of the identified regions if necessary.

We apply the source finding process to the 216\,MHz MWA mosaic, which has the best resolution and sensitivity and probes the part of the HII regions spectral profile where emission is expected to be brightest. Before applying the {\sc fellwalker} algorithm we first filter out the extended Galactic background emission by applying the spatial filtering algorithm {\sc findback}, which is also part of the {\sc starlink} project. This algorithm works by applying a three-stage filtering process. First, each pixel within a user specified box size is replaced with the minimum value. The box size specifies the minimum size of features that will remain in the image. The pixels in this minimum filtered image are then replaced by pixels with the maximum value. The final step is to replace each pixel by the mean value within the boxed region. A region size of $0.5 \times 0.5$\degree\ was chosen; this corresponds to the average size of the compact emission features seen in the image and resulted in the effective removal of the diffuse Galactic background emission. We did note a reduction in the flux density measurements for sources larger than the chosen region size in this filtered image. However, this does not effect the ability of the the {\sc fellwalker} algorithm to identify features. Finally, we create a 216\,MHz signal-to-noise image to account for variations in the image noise that would affect the thresholding source detection. 

\begin{figure}
\centering
\includegraphics[width=0.5\textwidth]{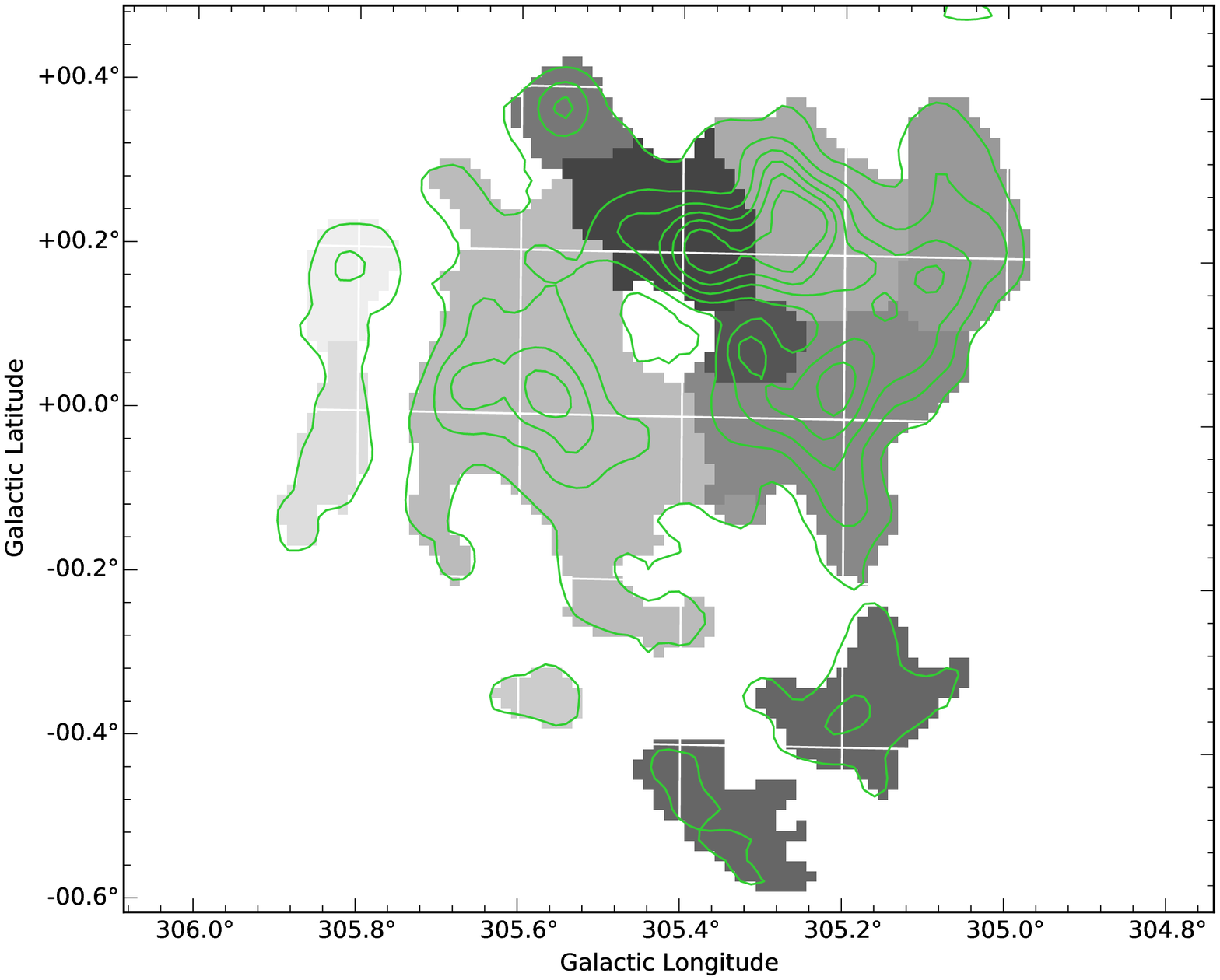} 
\includegraphics[width=0.5\textwidth]{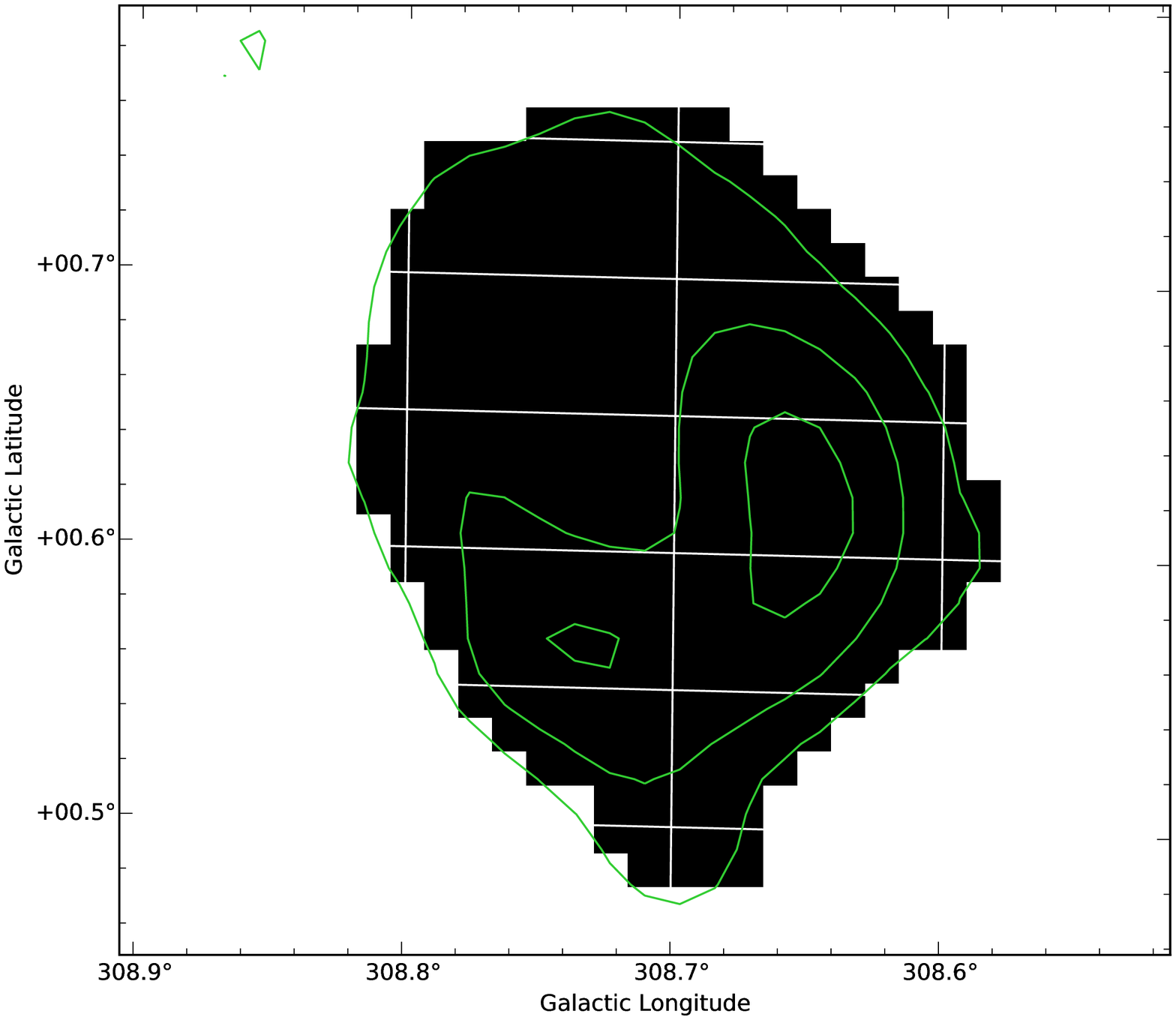} 
\caption{Here we show the output of our source finding process for a complex (top panel) and simple (bottom panel) source. The green contours are taken from our signal-to-noise 216\,MHz image and have a signal-to-noise level of 2, 10, 20, 30, 40, 50. In the top panel the different highlighted regions indicate where the algorithm has separated the region into distinct clumps. }
\label{im:template}
\end{figure}

The resultant template image was carefully checked by eye to ensure that the identified regions closely matched the emission features in the Galactic plane. We found that {\sc fellwalker} was able to accurately define emission features in the majority of cases. There were a few instances where the algorithm incorrectly features into multiple components or failed to properly trace very complex and blended emission features. In these cases we manually altered the template regions. An example of the template regions produced by {\sc fellwalker} can be seen in Fig.~\ref{im:template}.

	\subsection{Galactic Background estimation} \label{subsect:bkg}
\begin{figure}
\centering
\includegraphics[width=0.5\textwidth]{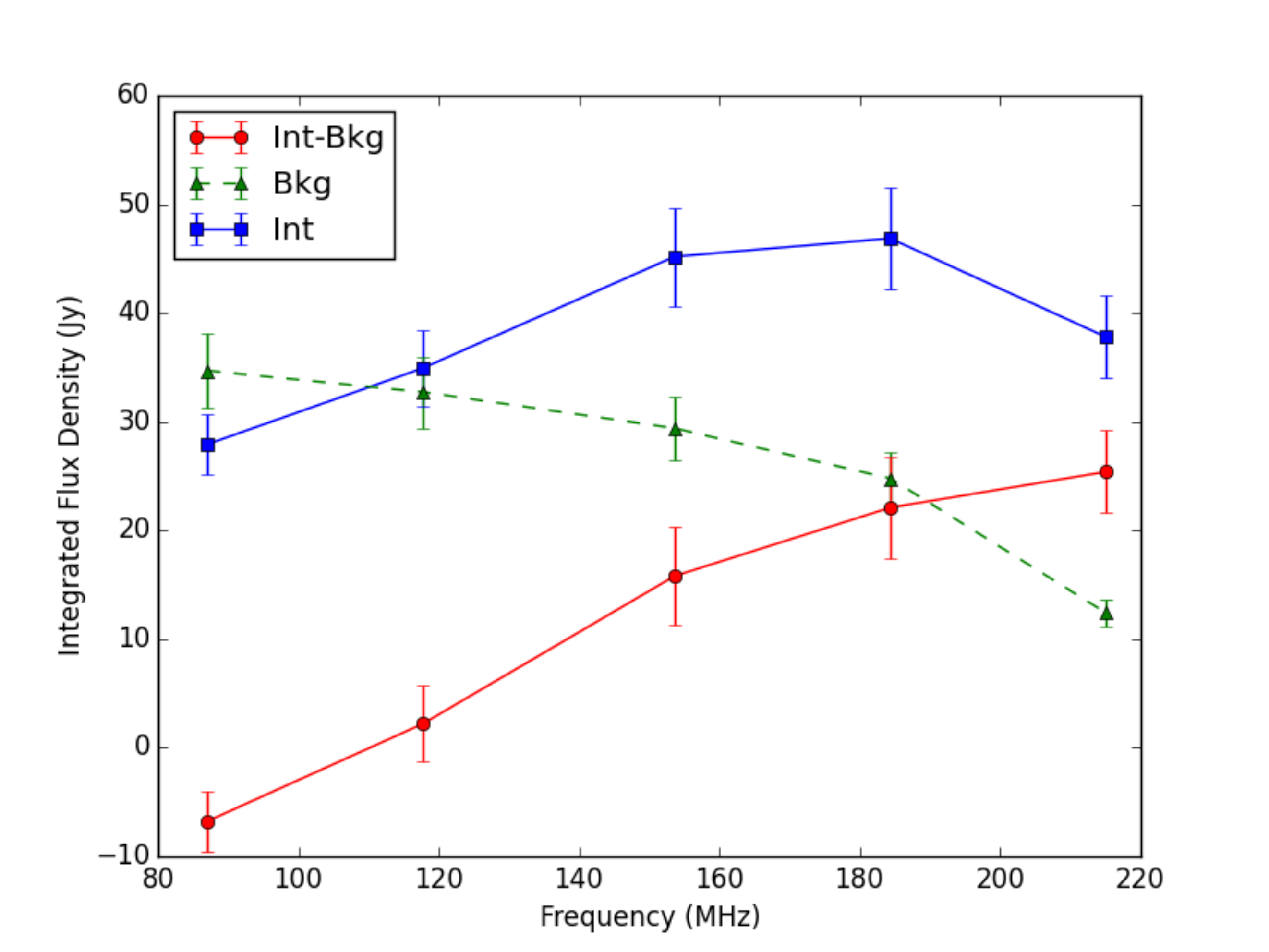} 
\includegraphics[width=0.55\textwidth]{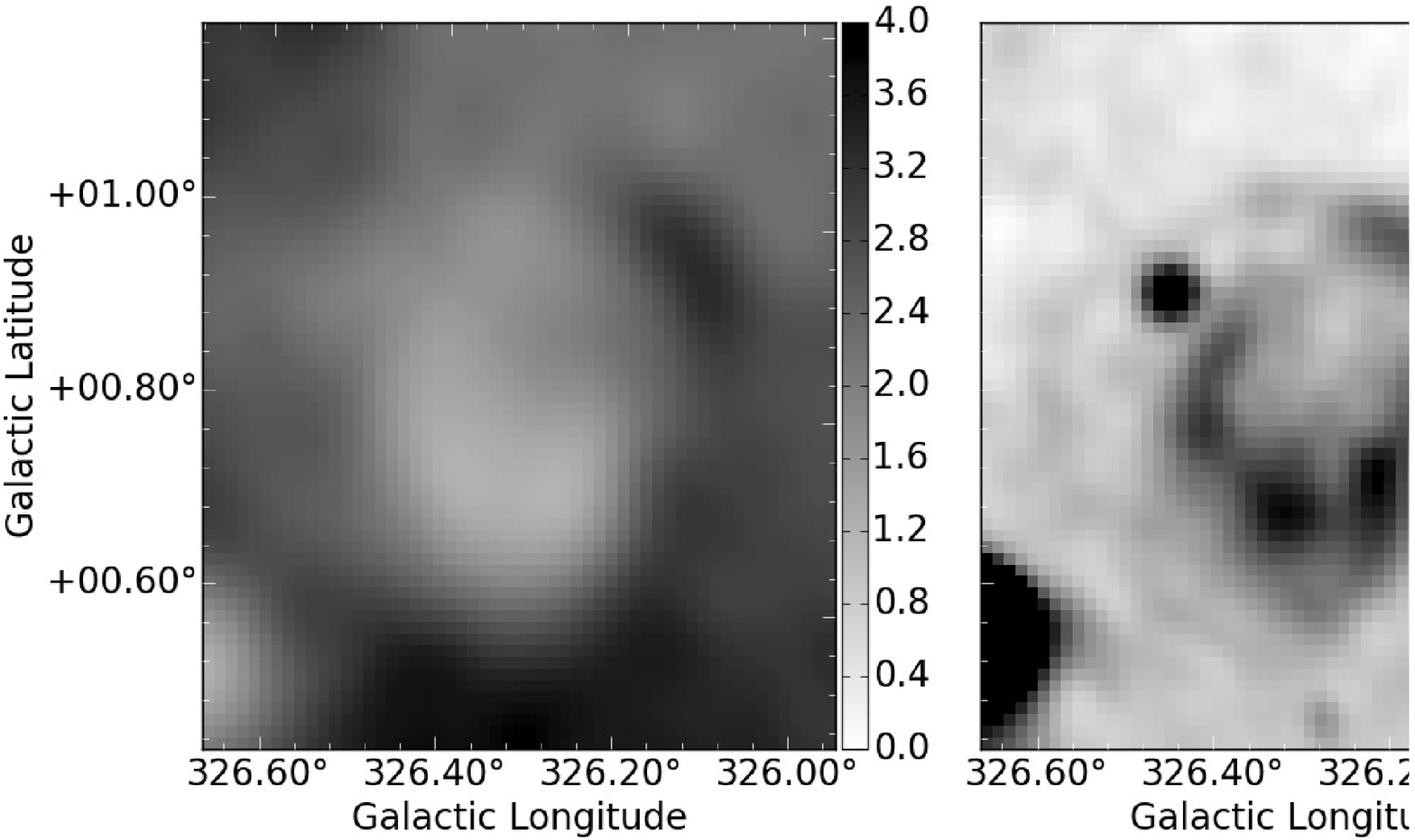} 
\caption{This plot shows the background subtraction for the G326.23+0.72 HII region. Top: blue squares indicate the integrated flux density; green triangles show the integrated background emission; and red circles show the integrated flux density with the background subtracted. Bottom: G326.23+0.72 at 88 (left panel) seen in absorption and 216\,MHz (right panel).}
\label{im:BKG}
\end{figure}

To estimate the integrated flux density and spectral index of HII regions in our MWA images we must first account for the large scale diffuse Galactic synchrotron background.  HII regions in our images are contaminated by the synchrotron background, which results in an increase in the measured integrated flux density at lower frequencies and a corresponding flattening of the HII region spectral index. The filtering process ({\sc findback}) used to eliminate the Galactic background in our source finding step was not suitable because it was found to decrease the flux density of larger sources. In an effort to characterise the Galactic background more accurately we make use of the {\sc bane} background estimation algorithm\footnote{https://github.com/PaulHancock/Aegean}. {\sc bane} applies a sliding boxcar filter over an image and provides an estimate of the background emission and noise. We found it difficult to obtain a good estimate of the Galactic background using {\sc Bane} alone. We were unable to produce an estimate of the background that did not contain residual emission from large and bright sources in the Galactic plane. We solved this problem by masking all emission features that were included in our initial {\sc fellwalker} template. We then run {\sc bane} to determine the background and use cubic interpolation to fill in the gaps where sources have been masked. Cubic interpolation was chosen over other methods such as linear or quadratic because it was found to more accurately recover the background emission. We then use this background image to subtract the estimated integrated background flux density from our raw HII region integrated flux densities. An example of the effect of background subtraction applied to the G326.23+0.72 HII region is shown in Fig.~\ref{im:BKG}.  We discuss the errors associated with this approach in Section~\ref{subsect:Uncertainties}. 

The background subtracted integrated flux density and spectral index is only applicable for HII regions that may be easily distinguished from the Galactic background synchrotron emission. In the cases where an HII region is blended with the Galactic synchrotron background we are unable to estimate the fraction of the observed flux that is due to the HII region or background. This occurs primarily in the 88 and 118\,MHz bands where the synchrotron background is brighter and the HII emission is fainter. If ignored then this effect would lead to very steep and non-physical spectral indices. We only fit the spectral index using frequency bands where the raw HII region integrated flux is at least two times greater than the integrated background. As an example the spectral index for the raw integrated flux density for G326.23+0.72 shown in Fig.~\ref{im:BKG} is $0.4\pm0.2$ (blue line), the background has a spectral index of $-0.9\pm0.3$ (green dashed line). When we subtract this background we find a spectral index for the HII region of $1.8\pm0.3$ (red line).

	\subsection{HII region identification} \label{subsect:ident}

\begin{figure}
\centering
\includegraphics[width=0.45\textwidth]{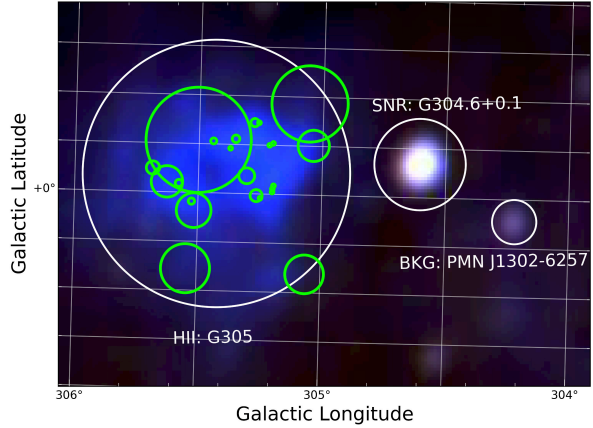} 
\caption{This three-colour image is constructed using the 88, 118, and 216\,MHz MWA images in red, green, and blue, respectively. This combination results in HII regions being seen as distinctive blue emission. HII regions identified as part of the G305 massive star forming complex by \cite{Anderson2014} are shown by green circles.}
\label{im:sources}
\vspace{-1.0em}
\end{figure}

The output of our source detection is an image with the features defined by an arbitrary number that can be used as a template to perform aperture photometry in our MWA images. The features identified by {\sc fellwalker} are comprised of primarily HII regions, SNRs, and background radio sources. We take advantage of the multiple bands provided by the MWA to identify HII regions by eye. We first construct a three-colour image using the 88, 118, and 216\,MHz bands in red, green, and blue, respectively (Fig.~\ref{im:sources}). Selecting this colour combination reveals optically thick HII regions by their distinct blue colour. Optically thick HII regions appear blue due to their steep positive spectral index of $\sim2.0$. SNRs, the Galactic background and background radio galaxies on the other hand appear to be red to white due to the steep negative spectral index of synchrotron emission $\sim-0.7$. To identify HII regions we first select all the features in our template that correspond to blue sources. We then verify that these sources have a steep spectral index characteristic of an optically thick HII region by extracting the integrated flux density using our source finding template and subtracting the corresponding Galactic background emission. We fit the SED across our MWA band and apply a spectral index selection criteria for HII regions, which requires the source to have a spectral index of $>0.0$. This limits our HII region sample to optically thick HII regions. However, we do not expect to detect optically thin HII regions given the low frequency, resolution, and sensitivity limits of the MWA. 

\begin{figure}
\centering
\includegraphics[width=0.5\textwidth]{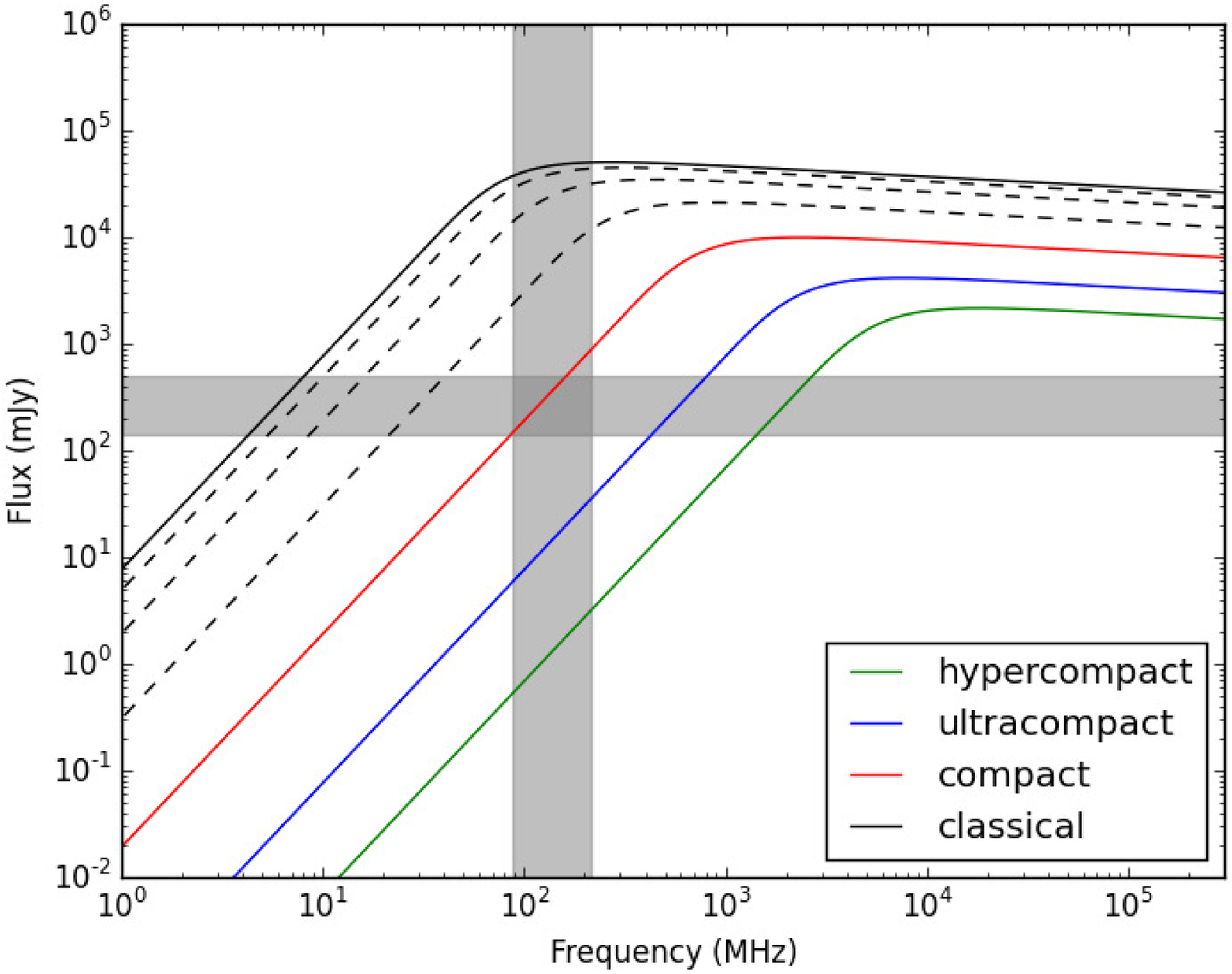} 
\includegraphics[width=0.5\textwidth]{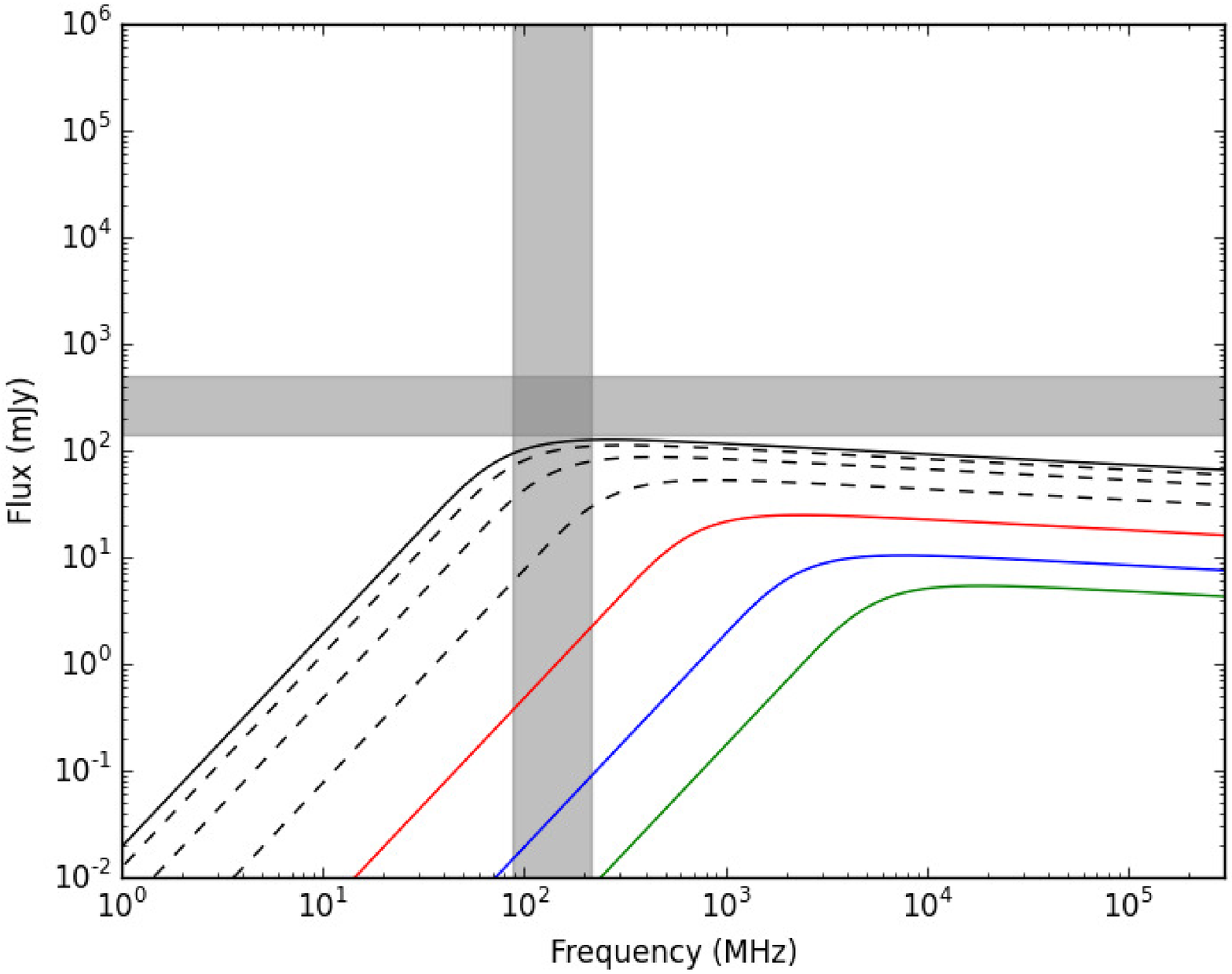} 
\caption{This plot shows the model SEDs for hypercompact (green), ultracompact (blue), compact (red) and classical (black) HII regions. The dashed black line shows the expected SED for HII regions that follow the relation $EM = 6.3\times10^5D^{-1.53\pm0.09}$ for diameters of 8, 5 and 2\,pc. The top panel shows the expected flux at 1\,kpc whilst the bottom panel shows the expected flux at 20\,kpc. The shaded region shows the 5-sigma sensitivity limit and bandwidth thresholds of our MWA images.}
\label{im:HII_model}
\end{figure}

As an HII region evolves its diameter ($s$) increases and the electron density ($n_{\rm e}$) and emission measure ($EM = \int n_{\rm e}^2 ds$) decreases. The relationship between emission measure and HII region diameter is $EM = 6.3\times10^5D^{-1.53\pm0.09}$ \citep{Garay1999}. This results in the frequency at which the radio continuum emission turns over from optically thin ($\alpha = -0.1$) to thick ($\alpha = 2.0$) to shift to lower frequencies as an HII region evolves. HII regions are commonly classified based on their size and emission measure into the following classes: hypercompact ($\lesssim0.03$\,pc, $\gtrsim 10^{9}$\,pc\,cm$^{-6}$), ultracompact ($\lesssim0.1$\,pc, $\gtrsim 10^{7}$\,pc\,cm$^{-6}$), compact ($\lesssim0.5$\,pc, $\gtrsim 10^{6}$\,pc\,cm$^{-6}$) and classical ($\sim10$\,pc, $\gtrsim 10^{4}$\,pc\,cm$^{-6}$). These classes represent a continuous distribution of HII region properties rather than discreet classes. Assuming a spherical, homogenous, isothermal HII region with an electron temperature of $10^4$\,K, we can plot the expected SED for these classes of HII regions using equations 1--4 of \cite{Mezger1967}. Figure~\ref{im:HII_model} shows that we are unlikely to detect any hypercompact or ultracompact HII regions given that their SED turns over in the GHz regime and then quickly drops off below our detection threshold. HII regions that we expect to detect will lie between the compact and classical classes. The SED of these types of HII region are expected to be either optically thin or within the turnover regime where the optical depth is unity ($\tau=1$). This figure also demonstrates that we are likely to miss HII regions at large Galactic distances ($\sim 20$\,kpc) due to the sensitivity of our observations.

To demonstrate the effectiveness of this method of selecting optically thick HII regions we present a small subregion in Fig.~\ref{im:sources}. The massive star forming complex known as G305 is comprised of a number of HII regions \citep{Clark2004,Hindson2013} and can be clearly seen as bright blue emission whilst the SNR G304.6+0.1 \citep{Green2014} and background source PMN\,J1302-6257 appears bright white. We take advantage of these spectral properties to detect SNRs in the Galactic plane and present the results in \cite{Hollitt2015}. 

The resolution of the MWA limits us to identifying HII regions with sizes $>1$ to $>31$\,pc depending on distance (1--20\,kpc). Given the frequency regime and sensitivity of the MWA we can expect to detect optically thick HII regions from the compact to classical stage with sizes of $\gtrsim 0.5$ to $>10$\,pc. Massive star formation occurs in large star forming complexes which results in multiple epochs of HII regions, with a range of sizes, in close proximity \cite{Murray2010}. It is therefore unsurprising that we see evidence of blended HII regions present in massive star forming complexes such as in the case of G305. In Fig.~\ref{im:sources} we overlay the HII regions identified by the high resolution (6--12\arcsec) detected HII regions of \cite{Anderson2014}, which was compiled using the Wide-Field Infrared Survey Explorer (WISE: \citealt{Wright2010}) at MIR wavelengths from 3.4 to $22\mu$m and ancillary data. G305 is clearly associated with many more HII regions than we are able to resolve.

	\subsection{Uncertainties}
	\label{subsect:Uncertainties}
\begin{figure}
\centering
\includegraphics[width=0.5\textwidth]{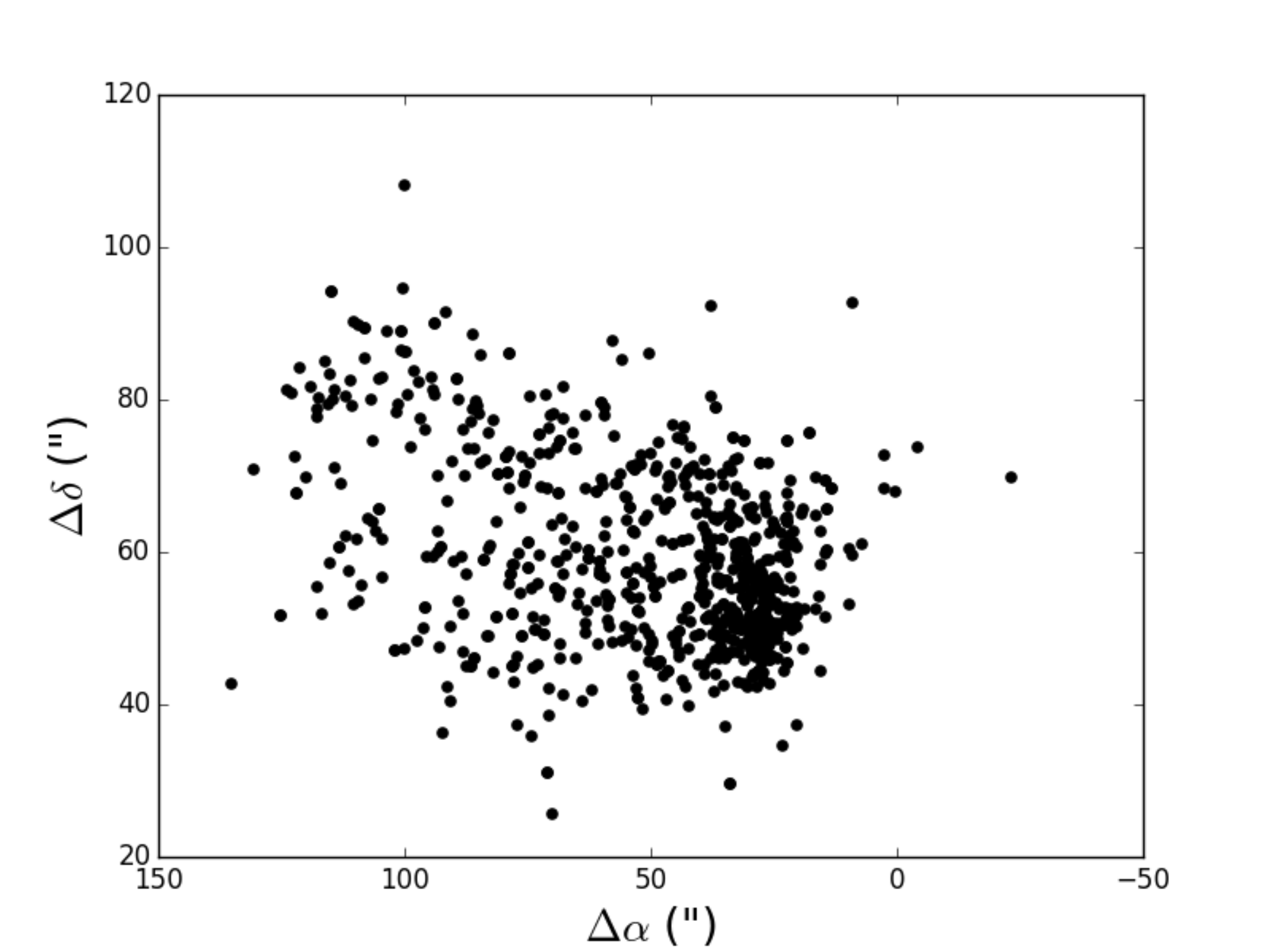} 
\caption{Plot of the offset (MWA position minus MRC position) in right ascension ($\Delta \alpha$) and declination ($\Delta \delta$) in units of arc-seconds for 792 bright point sources found within $\pm 15$\degree\ of the Galactic plane in our 154\,MHz MWA image.}
\label{im:offset}
\end{figure}

To determine the positional uncertainty in each of our MWA images we compare the peak flux density position of MWA point sources, compiled using {\sc Aegean}, to those in the MRC survey within $\pm 15$\degree\ of the Galactic plane. We find a persistent positional offset between our MWA sources and the matched MRC sources in both right ascension and declination (Fig.~\ref{im:offset}). We find an average offset for each increasing band from 88 to 216\,MHz of 102, 89, 80, 69, and 60\arcsec, respectively. These offsets are smaller than the synthesised beam in each band and correspond to an offset of approximately a third of a beam. The source of this offset lies in the $w$-term and ionospheric effects. Given the large synthesised beam and extended nature of the majority of our HII region sample we find that this offset is an acceptable level of accuracy.

Centaurus A (Cen\,A) is responsible for introducing significant contamination resulting in an approximately two-fold increase in the local noise between Galactic longitudes of $206\, <\, l\, <\, 307$\degree. Attempts were made to peel Cen\,A from the image using standard peeling techniques (e.g., \citealt{Mitchell2008}). However, due to the complex nature and extreme brightness of Cen\,A (peak flux density of 456\,Jy at 154\,MHz) we were unable to satisfactorily peel the source. The flux density estimates in our catalogue within this region reflect the increase in the local noise caused by Cen\,A.

To determine the accuracy of our background subtraction method we inject a Gaussian shaped model source with a randomly selected flux density, size, and position into each of our MWA images. In each band we choose a flux density range that starts at 1\,Jy and upper limits of 40, 30, 20, 15 and 10\,Jy for each increasing band starting from 88\,MHz. The size of the Gaussian sources range from 0.1 to 1.0\degree. We compare the integrated background flux density in our MWA image before injecting the source to the background estimated after applying our background estimation approach described in Section~\ref{subsect:bkg}. We carry out this procedure 1000 times and find that on average our estimated background is accurate to approximately 20\% for each band.

The uncertainty for the flux densities are calculated using the uncertainty derived from the flux density scale, local noise uncertainty, and the error associated with the background estimation where appropriate combined in quadrature.

Optically thick SNRs and absorption processes could potentially contaminate our selection method, which is based on a source having a steep positive spectral index. The spectral index of SNRs can increase due to a number of processes including: synchrotron self-absorption; intrinsic free-free self-absorption due to thermal ejecta; and extrinsic free-free absorption by a foreground source. Synchrotron self-absorption is expected to occur only in very bright and very compact SNRs and results in spectral indices between $\alpha = 2.0$ and $2.5$ \citep{Longair2011,Ginzburg1969}. However, for a typical SNR with a solid angle of 1 arcminute$^2$, magnetic field of 10$\mu$G and flux density of 100\,Jy at 150\,MHz synchrotron self-absorption would occur at 1\,MHz. It is therefore highly unlikely that such a process would contaminate our results and we can disregard this mechanism. In the case of intrinsic free-free self-absorption, thermal ejecta associated with young SNRs (on the order of $10^3$\,yrs) may cause a flattening of the SED towards the centre of SNRs. For example, the resolved spectral index of the SNR Cassiopeia\,A has been found to flatten to $-0.35$ between 330 and 74\,MHz \citep{Brogan2005,Delaney2014} due to this effect. The spectra of such SNRs is still dominated by the synchrotron process and we do not expect such young SNRs to be resolved by the MWA except in the case of nearby and obvious SNRs such as Vela. We note that the spectral index of the Vela SNR, located at $l=263.9$, $b=-03.3$, ranges from $-1.3$ towards the edge to $0.2$ towards the centre with an integrated spectral index of $-0.3\pm0.1$ between 88 and 216\,MHz. This is in good agreement with previous findings where the spectral integrated index is approximately $-0.39\pm0.03$ between 30 and 8400\,MHz \citep{Alvarez2001}. The shallow spectral index of Vela does not satisfy our selection criteria ($\alpha > 0.0$) but does result in a somewhat blue colour in our three-colour image towards the centre of Vela. Finally, it has been suggested that free-free absorption along the line of sight is responsible for a turnover in the SED of $\sim 2/3$ of Galactic SNRs below 100\,MHz \citep{Kassim1989,Lacey2001}. The foreground sources responsible may be due to low-density ($n_{\rm e}\sim1$--$10$\,cm$^{-3}$), intermediate temperature ($T\sim5000$\,K) ionised thermal gas or HII regions. We would only expect such SNRs to turnover below 100\,MHz. Given the frequency bands of the MWA this would result in only the 88\,MHz band being affected which would not lead to a significant change in our spectral index measurements.

Both mechanisms of absorption in SNRs by the free-free mechanism are observed at frequencies below 100\,MHz. In comparison HII regions with typical physical properties, see Section~\ref{subsect:thick}, are expected to be optically thick between approximately 10 and 400\,MHz and optically thin above 400\,MHz. Given the resolution and frequency band coverage of the MWA we do not expect to detect SNRs that could be mistaken as HII regions. In fact the frequency coverage of the MWA provides us with the unique opportunity to identify HII regions that are projected against SNRs. To check for any contamination we compare our catalogue of HII regions to the catalogue of SNRs compiled by \cite{Green2014}. We find that none of our HII regions are coincident with known SNRs presented in the \cite{Green2014} catalogue. 

The low resolution of these observations results in the blending of discrete HII regions. For sources that are part of the same complex this leads to an averaging of the emission across multiple HII regions in the complex. For cases where HII regions lie at different distances but along the same line of sight HII regions that are not physically associated will be blended together. 

\section{Results}
\label{sec:results}

\begin{landscape}
\begin{figure}
\begin{center}
\includegraphics[width=1.3\textwidth,trim={0 0 0 20}]{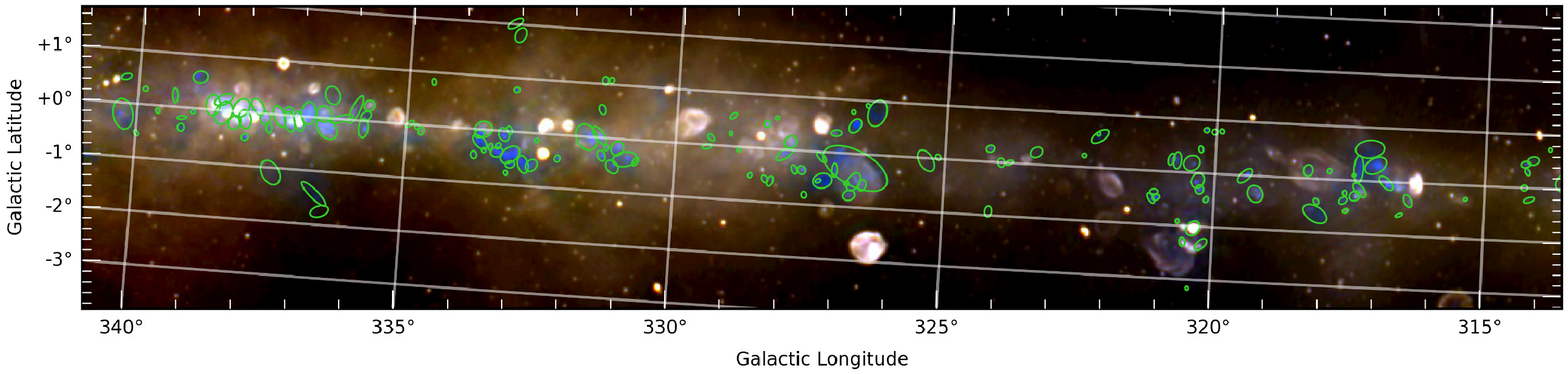} \includegraphics[width=1.3\textwidth]{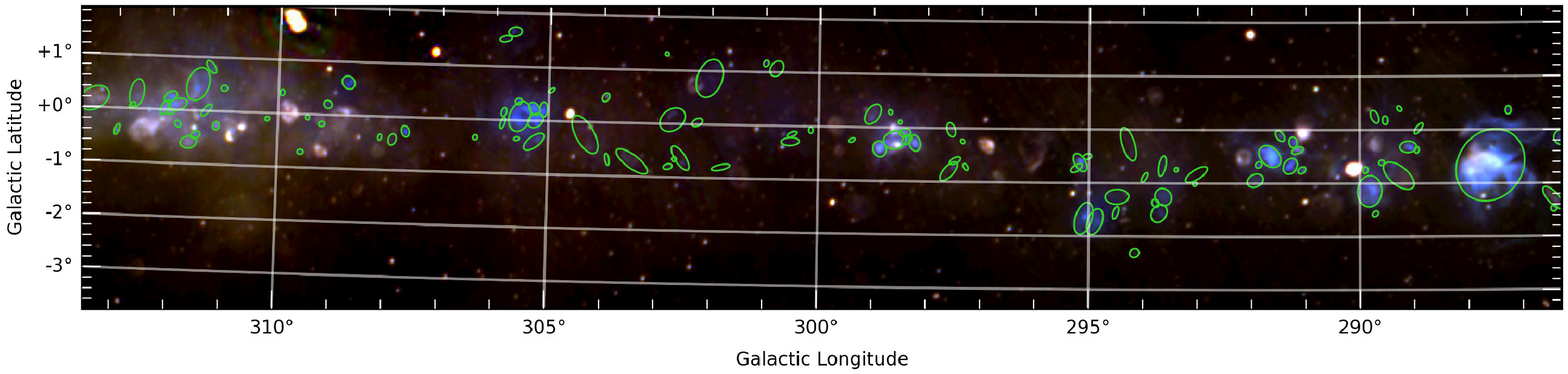} 
\includegraphics[width=1.3\textwidth]{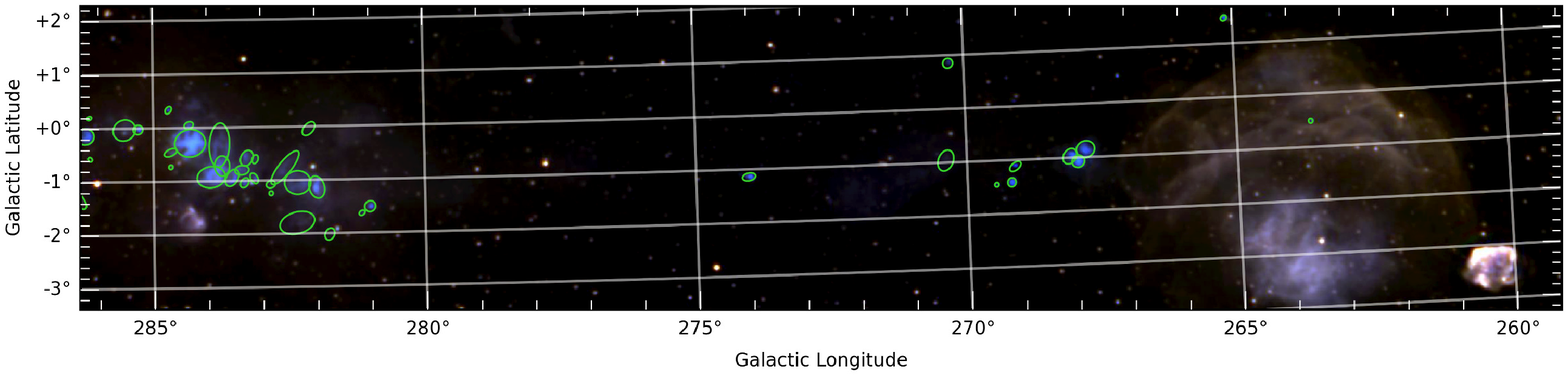} 
\caption{Three-colour image generated using our 88, 118, and 216\,MHz MWA images in red, green, and blue, respectively. At these low frequencies the thermal free-free emission of HII regions leads to a distinctly steep spectral index of $\sim 2.0$ compared to the $-0.8$ to $-0.5$ spectral index of the Galactic background and supernova remnants, respectively. Using this colour scheme results in HII regions appearing blue, whilst other features appear white to red. HII regions detected in our source-finding step are indicated by green ellipses.}
\label{im:main}
\end{center}
\end{figure}
\end{landscape}

We find a total of 302 HII regions, which are shown in Fig.~\ref{im:main}. We present a small section of the resultant catalogue in Table~\ref{tab:main}. This table reports the background subtracted integrated flux density and associated spectral index. A machine readable version of the complete catalogue, which includes the raw and background integrated flux densities can be found online via Vizier\footnote{http://vizier.u-strasbg.fr/}.

We identify 45 optically thick HII regions that can be seen as absorption features against the diffuse Galactic background synchrotron emission. Theses HII have an integrated flux density at 88\,MHz that is lower than the associated Galactic background. An example of such an HII region can be seen in the bottom left panel of Fig.~\ref{im:BKG}. These HII regions are concentrated towards the Galactic centre, between 314 and 340\degree, where the diffuse Galactic background is brightest.

\begin{landscape}

\begin{table}
 \caption{The first 50 of 302 HII regions identified in this study with increasing distance from the Galactic centre. The integrated flux density measurements have had the background component subtracted. Due to space limitations we present only the 88, 185, and 215\,MHz results here and refer the reader to the full version of the catalogue which may be found online. This full version also contains the integrated flux density of MGPS and SGPS sources where available.}
\scriptsize
\begin{tabular}{l|cccc|ccc|ccc|c}
 \hline 
Name & \multicolumn{2}{c}{Position (Gal)} & \multicolumn{2}{c}{Size (\degree)} & \multicolumn{3}{c}{Peak Flux Density (Jy/beam)} & \multicolumn{3}{c}{Integrated Flux Density (Jy)} & Spectral Index \\
     & $(l)$ & $(b)$ & $(x)$ & $(y)$ & 88\,MHz & 185\,MHz & 216\,MHz & 88\,MHz & 185\,MHz & 216\,MHz & ($\alpha$)\\
 			
 	\hline
G263.62-0.52 & 263.62 & -0.52 & 0.04 & 0.04 & $ 3.4 \pm 0.3 $ & $ 1.9 \pm 0.2 $ & $ 2.1 \pm 0.2 $ & $ 0.4 \pm 0.1 $ & $ 1.1 \pm 0.3 $ & $ 2.2 \pm 0.5 $ & $ - $ \\
G265.16+1.46 & 265.16 & 1.46 & 0.06 & 0.05 & $ 2.5 \pm 0.3 $ & $ 3.6 \pm 0.4 $ & $ 5.4 \pm 0.5 $ & $ 3.3 \pm 0.8 $ & $ 9.4 \pm 2.1 $ & $ 17.1 \pm 3.9 $ & $ 1.8 \pm 0.3 $ \\
G267.82-0.88 & 267.82 & -0.88 & 0.17 & 0.15 & $ 3.0 \pm 0.3 $ & $ 3.0 \pm 0.3 $ & $ 3.3 \pm 0.3 $ & $ 14.9 \pm 3.5 $ & $ 32.2 \pm 7.3 $ & $ 52.7 \pm 11.9 $ & $ 1.4 \pm 0.3 $ \\
G267.95-1.07 & 267.95 & -1.07 & 0.12 & 0.11 & $ 3.5 \pm 0.4 $ & $ 5.2 \pm 0.5 $ & $ 8.1 \pm 0.8 $ & $ 11.9 \pm 2.7 $ & $ 34.7 \pm 7.8 $ & $ 62.5 \pm 14.0 $ & $ 1.8 \pm 0.3 $ \\
G268.05-0.96 & 268.05 & -0.96 & 0.16 & 0.12 & $ 3.2 \pm 0.3 $ & $ 3.4 \pm 0.3 $ & $ 3.8 \pm 0.4 $ & $ 9.6 \pm 2.2 $ & $ 22.7 \pm 5.1 $ & $ 34.4 \pm 7.8 $ & $ 1.4 \pm 0.3 $ \\
G269.12-1.13 & 269.12 & -1.13 & 0.12 & 0.07 & $ 1.6 \pm 0.2 $ & $ 2.3 \pm 0.2 $ & $ 3.7 \pm 0.4 $ & $ 1.1 \pm 0.4 $ & $ 3.6 \pm 0.9 $ & $ 7.8 \pm 1.8 $ & $ 2.2 \pm 0.4 $ \\
G269.19-1.46 & 269.19 & -1.46 & 0.08 & 0.08 & $ 2.9 \pm 0.3 $ & $ 3.2 \pm 0.3 $ & $ 3.2 \pm 0.3 $ & $ 4.0 \pm 0.9 $ & $ 11.3 \pm 2.6 $ & $ 17.9 \pm 4.0 $ & $ 1.6 \pm 0.3 $ \\
G269.47-1.46 & 269.47 & -1.46 & 0.04 & 0.04 & $ 1.2 \pm 0.1 $ & $ 0.9 \pm 0.1 $ & $ 1.2 \pm 0.1 $ & $ 0.4 \pm 0.1 $ & $ 0.8 \pm 0.2 $ & $ 1.5 \pm 0.4 $ & $ - $ \\
G270.27+0.86 & 270.27 & 0.86 & 0.09 & 0.09 & $ 1.7 \pm 0.2 $ & $ 1.4 \pm 0.1 $ & $ 1.4 \pm 0.1 $ & $ 2.8 \pm 0.7 $ & $ 4.5 \pm 1.1 $ & $ 6.2 \pm 1.5 $ & $ 0.9 \pm 0.3 $ \\
G270.41-0.95 & 270.41 & -0.95 & 0.20 & 0.14 & $ 1.8 \pm 0.2 $ & $ 0.6 \pm 0.1 $ & $ 0.5 \pm 0.1 $ & $ 4.3 \pm 1.1 $ & $ 4.8 \pm 1.2 $ & $ 7.0 \pm 1.8 $ & $ 0.5 \pm 0.4 $ \\
G274.01-1.15 & 274.01 & -1.15 & 0.12 & 0.07 & $ 3.5 \pm 0.4 $ & $ 4.3 \pm 0.4 $ & $ 6.2 \pm 0.6 $ & $ 6.6 \pm 1.5 $ & $ 13.7 \pm 3.1 $ & $ 22.9 \pm 5.2 $ & $ 1.4 \pm 0.3 $ \\
G281.00-1.52 & 281.00 & -1.52 & 0.11 & 0.10 & $ 4.4 \pm 0.4 $ & $ 3.5 \pm 0.4 $ & $ 3.4 \pm 0.3 $ & $ 6.4 \pm 1.5 $ & $ 9.7 \pm 2.2 $ & $ 13.8 \pm 3.2 $ & $ 0.8 \pm 0.3 $ \\
G281.18-1.65 & 281.18 & -1.65 & 0.06 & 0.04 & $ 2.7 \pm 0.3 $ & $ 1.4 \pm 0.1 $ & $ 1.6 \pm 0.2 $ & $ 0.7 \pm 0.2 $ & $ 1.0 \pm 0.3 $ & $ 1.8 \pm 0.4 $ & $ - $ \\
G281.73-2.03 & 281.73 & -2.03 & 0.12 & 0.08 & $ 3.5 \pm 0.4 $ & $ 1.0 \pm 0.1 $ & $ 0.6 \pm 0.1 $ & $ 1.5 \pm 0.4 $ & $ 2.1 \pm 0.5 $ & $ 2.6 \pm 0.8 $ & $ - $ \\
G282.03-1.17 & 282.03 & -1.17 & 0.21 & 0.14 & $ 6.3 \pm 0.6 $ & $ 5.6 \pm 0.6 $ & $ 7.4 \pm 0.7 $ & $ 21.6 \pm 5.0 $ & $ 35.4 \pm 8.0 $ & $ 56.1 \pm 12.7 $ & $ 1.0 \pm 0.3 $ \\
G282.13-0.09 & 282.13 & -0.09 & 0.15 & 0.09 & $ 2.0 \pm 0.2 $ & $ 0.7 \pm 0.1 $ & $ 0.5 \pm 0.1 $ & $ 1.0 \pm 0.4 $ & $ 2.0 \pm 0.6 $ & $ 2.3 \pm 0.8 $ & $ - $ \\
G282.23-1.08 & 282.23 & -1.08 & 0.24 & 0.22 & $ 4.5 \pm 0.5 $ & $ 2.1 \pm 0.2 $ & $ 1.7 \pm 0.2 $ & $ 20.1 \pm 4.7 $ & $ 32.4 \pm 7.4 $ & $ 46.6 \pm 10.7 $ & $ 0.9 \pm 0.3 $ \\
G282.34-1.83 & 282.34 & -1.83 & 0.33 & 0.20 & $ 3.0 \pm 0.3 $ & $ 1.2 \pm 0.1 $ & $ 0.8 \pm 0.1 $ & $ 9.2 \pm 2.4 $ & $ 10.7 \pm 2.7 $ & $ 14.5 \pm 3.9 $ & $ 0.5 \pm 0.4 $ \\
G282.63-0.85 & 282.63 & -0.85 & 0.39 & 0.12 & $ 3.9 \pm 0.4 $ & $ 1.6 \pm 0.2 $ & $ 1.3 \pm 0.1 $ & $ 11.7 \pm 2.8 $ & $ 16.1 \pm 3.8 $ & $ 22.9 \pm 5.5 $ & $ 0.7 \pm 0.3 $ \\
G282.84-1.25 & 282.84 & -1.25 & 0.04 & 0.04 & $ 2.6 \pm 0.3 $ & $ 1.1 \pm 0.1 $ & $ 1.1 \pm 0.1 $ & $ 0.2 \pm 0.1 $ & $ 0.6 \pm 0.1 $ & $ 0.9 \pm 0.2 $ & $ - $ \\
G282.87-1.06 & 282.87 & -1.06 & 0.09 & 0.07 & $ 3.1 \pm 0.3 $ & $ 1.1 \pm 0.1 $ & $ 0.8 \pm 0.1 $ & $ 1.2 \pm 0.3 $ & $ 1.8 \pm 0.4 $ & $ 2.3 \pm 0.6 $ & $ - $ \\
G283.12-0.61 & 283.12 & -0.61 & 0.08 & 0.06 & $ 3.5 \pm 0.4 $ & $ 1.2 \pm 0.1 $ & $ 0.9 \pm 0.1 $ & $ 1.2 \pm 0.3 $ & $ 1.9 \pm 0.5 $ & $ 3.1 \pm 0.7 $ & $ - $ \\
G283.13-0.99 & 283.13 & -0.99 & 0.10 & 0.06 & $ 6.0 \pm 0.6 $ & $ 2.2 \pm 0.2 $ & $ 2.3 \pm 0.2 $ & $ 3.4 \pm 0.8 $ & $ 5.1 \pm 1.2 $ & $ 7.9 \pm 1.8 $ & $ 0.9 \pm 0.3 $ \\
G283.32-0.55 & 283.32 & -0.55 & 0.15 & 0.11 & $ 4.5 \pm 0.5 $ & $ 1.9 \pm 0.2 $ & $ 1.8 \pm 0.2 $ & $ 6.9 \pm 1.6 $ & $ 12.3 \pm 2.8 $ & $ 17.5 \pm 4.0 $ & $ 1.0 \pm 0.3 $ \\
G283.33-1.05 & 283.33 & -1.05 & 0.10 & 0.07 & $ 6.0 \pm 0.6 $ & $ 2.5 \pm 0.3 $ & $ 2.2 \pm 0.2 $ & $ 4.0 \pm 0.9 $ & $ 6.5 \pm 1.5 $ & $ 9.6 \pm 2.2 $ & $ 0.9 \pm 0.3 $ \\
G283.44-0.80 & 283.44 & -0.80 & 0.13 & 0.09 & $ 4.9 \pm 0.5 $ & $ 1.9 \pm 0.2 $ & $ 1.9 \pm 0.2 $ & $ 5.6 \pm 1.3 $ & $ 8.9 \pm 2.0 $ & $ 13.5 \pm 3.1 $ & $ 0.9 \pm 0.3 $ \\
G283.54-1.02 & 283.54 & -1.02 & 0.17 & 0.11 & $ 5.2 \pm 0.5 $ & $ 2.7 \pm 0.3 $ & $ 2.9 \pm 0.3 $ & $ 12.0 \pm 2.7 $ & $ 21.2 \pm 4.8 $ & $ 30.8 \pm 6.9 $ & $ 1.0 \pm 0.3 $ \\
G283.79-0.78 & 283.79 & -0.78 & 0.19 & 0.14 & $ 5.5 \pm 0.6 $ & $ 2.9 \pm 0.3 $ & $ 2.9 \pm 0.3 $ & $ 13.5 \pm 3.1 $ & $ 24.6 \pm 5.5 $ & $ 35.8 \pm 8.1 $ & $ 1.0 \pm 0.3 $ \\
G283.81-0.63 & 283.81 & -0.63 & 0.41 & 0.19 & $ 4.5 \pm 0.5 $ & $ 2.1 \pm 0.2 $ & $ 2.0 \pm 0.2 $ & $ 17.3 \pm 4.1 $ & $ 28.9 \pm 6.6 $ & $ 42.8 \pm 9.8 $ & $ 0.9 \pm 0.3 $ \\
G283.98-0.95 & 283.98 & -0.95 & 0.26 & 0.20 & $ 5.9 \pm 0.6 $ & $ 3.2 \pm 0.3 $ & $ 3.4 \pm 0.3 $ & $ 38.1 \pm 8.6 $ & $ 71.9 \pm 16.2 $ & $ 102.8 \pm 23.2 $ & $ 1.1 \pm 0.3 $ \\
G284.32-0.35 & 284.32 & -0.35 & 0.29 & 0.25 & $ 6.1 \pm 0.6 $ & $ 6.1 \pm 0.6 $ & $ 10.5 \pm 1.0 $ & $ 76.1 \pm 17.3 $ & $ 186.5 \pm 41.9 $ & $ 322.1 \pm 72.2 $ & $ 1.5 \pm 0.3 $ \\
G284.36+0.03 & 284.36 & 0.03 & 0.09 & 0.06 & $ 4.0 \pm 0.4 $ & $ 2.0 \pm 0.2 $ & $ 1.9 \pm 0.2 $ & $ 2.9 \pm 0.7 $ & $ 5.7 \pm 1.3 $ & $ 8.0 \pm 1.8 $ & $ 1.1 \pm 0.3 $ \\
G284.64-0.50 & 284.64 & -0.50 & 0.13 & 0.06 & $ 4.4 \pm 0.4 $ & $ 2.0 \pm 0.2 $ & $ 1.8 \pm 0.2 $ & $ 4.2 \pm 1.0 $ & $ 6.1 \pm 1.4 $ & $ 6.9 \pm 1.6 $ & $ 0.6 \pm 0.3 $ \\
G284.68-0.73 & 284.68 & -0.73 & 0.04 & 0.04 & $ 4.7 \pm 0.5 $ & $ 1.4 \pm 0.1 $ & $ 1.2 \pm 0.1 $ & $ 0.9 \pm 0.2 $ & $ 0.9 \pm 0.2 $ & $ 1.0 \pm 0.3 $ & $ - $ \\
G284.73+0.32 & 284.73 & 0.32 & 0.07 & 0.05 & $ 3.0 \pm 0.3 $ & $ 2.0 \pm 0.2 $ & $ 2.7 \pm 0.3 $ & $ 1.1 \pm 0.3 $ & $ 2.5 \pm 0.6 $ & $ 4.6 \pm 1.1 $ & $ 1.8 \pm 1.0 $ \\
G285.27-0.04 & 285.27 & -0.04 & 0.09 & 0.08 & $ 3.7 \pm 0.4 $ & $ 3.5 \pm 0.3 $ & $ 5.5 \pm 0.6 $ & $ 4.4 \pm 1.1 $ & $ 8.5 \pm 2.0 $ & $ 15.3 \pm 3.5 $ & $ 1.3 \pm 0.3 $ \\
G285.54-0.10 & 285.54 & -0.10 & 0.21 & 0.19 & $ 4.6 \pm 0.5 $ & $ 1.9 \pm 0.2 $ & $ 1.6 \pm 0.2 $ & $ 10.1 \pm 2.5 $ & $ 13.3 \pm 3.3 $ & $ 19.6 \pm 4.8 $ & $ 0.6 \pm 0.4 $ \\
G286.17+0.19 & 286.17 & 0.19 & 0.04 & 0.03 & $ 2.5 \pm 0.3 $ & $ 0.7 \pm 0.1 $ & $ 0.6 \pm 0.1 $ & $ 0.3 \pm 0.1 $ & $ 0.3 \pm 0.1 $ & $ 0.7 \pm 0.2 $ & $ - $ \\
G286.17-0.58 & 286.17 & -0.58 & 0.04 & 0.03 & $ 2.4 \pm 0.3 $ & $ 0.7 \pm 0.1 $ & $ 0.5 \pm 0.1 $ & $ 0.2 \pm 0.1 $ & $ 0.3 \pm 0.1 $ & $ 0.4 \pm 0.1 $ & $ - $ \\
G286.19-0.16 & 286.19 & -0.16 & 0.16 & 0.15 & $ 4.1 \pm 0.4 $ & $ 3.3 \pm 0.3 $ & $ 3.2 \pm 0.3 $ & $ 9.8 \pm 2.4 $ & $ 28.0 \pm 6.4 $ & $ 41.5 \pm 9.4 $ & $ 1.6 \pm 0.3 $ \\
G286.39-1.35 & 286.39 & -1.35 & 0.26 & 0.10 & $ 5.2 \pm 0.5 $ & $ 1.9 \pm 0.2 $ & $ 1.6 \pm 0.2 $ & $ 14.8 \pm 3.4 $ & $ 13.4 \pm 3.2 $ & $ 16.7 \pm 4.0 $ & $ - $ \\
G286.43-1.48 & 286.43 & -1.48 & 0.06 & 0.05 & $ 3.7 \pm 0.4 $ & $ 1.6 \pm 0.2 $ & $ 1.3 \pm 0.1 $ & $ 1.4 \pm 0.3 $ & $ 2.3 \pm 0.5 $ & $ 3.0 \pm 0.7 $ & $ - $ \\
G287.25+0.36 & 287.25 & 0.36 & 0.08 & 0.05 & $ 2.3 \pm 0.2 $ & $ 1.3 \pm 0.1 $ & $ 1.3 \pm 0.1 $ & $ 1.0 \pm 0.3 $ & $ 2.1 \pm 0.5 $ & $ 3.8 \pm 0.9 $ & $ 1.8 \pm 1.0 $ \\
G287.41-0.64 & 287.41 & -0.64 & 0.70 & 0.60 & $ 13.8 \pm 1.4 $ & $ 8.3 \pm 0.8 $ & $ 10.2 \pm 1.0 $ & $ 408.1 \pm 92.4 $ & $ 814.9 \pm 183.0 $ & $ 1214.7 \pm 272.7 $ & $ 1.2 \pm 0.3 $ \\
G288.94-0.02 & 288.94 & -0.02 & 0.12 & 0.05 & $ 2.5 \pm 0.3 $ & $ 0.8 \pm 0.1 $ & $ 0.7 \pm 0.1 $ & $ 0.4 \pm 0.2 $ & $ 0.4 \pm 0.2 $ & $ 1.0 \pm 0.3 $ & $ - $ \\
G288.94-0.40 & 288.94 & -0.40 & 0.06 & 0.04 & $ 4.1 \pm 0.4 $ & $ 1.6 \pm 0.2 $ & $ 1.2 \pm 0.1 $ & $ 1.0 \pm 0.2 $ & $ 1.5 \pm 0.4 $ & $ 1.7 \pm 0.4 $ & $ - $ \\
G289.08-0.36 & 289.08 & -0.36 & 0.16 & 0.10 & $ 5.5 \pm 0.6 $ & $ 3.7 \pm 0.4 $ & $ 4.1 \pm 0.4 $ & $ 10.3 \pm 2.4 $ & $ 15.7 \pm 3.6 $ & $ 24.3 \pm 5.6 $ & $ 0.9 \pm 0.3 $ \\
G289.25+0.37 & 289.25 & 0.37 & 0.06 & 0.04 & $ 2.3 \pm 0.3 $ & $ 0.6 \pm 0.1 $ & $ 0.4 \pm 0.1 $ & $ 0.2 \pm 0.1 $ & $ 0.2 \pm 0.1 $ & $ 0.4 \pm 0.2 $ & $ - $ \\
G289.40-0.68 & 289.40 & -0.68 & 0.34 & 0.17 & $ 5.6 \pm 0.6 $ & $ 1.7 \pm 0.2 $ & $ 1.4 \pm 0.1 $ & $ 17.3 \pm 4.0 $ & $ 14.9 \pm 3.6 $ & $ 20.8 \pm 5.0 $ & $ - $ \\
G289.51+0.14 & 289.51 & 0.14 & 0.08 & 0.05 & $ 3.1 \pm 0.3 $ & $ 1.4 \pm 0.1 $ & $ 1.1 \pm 0.1 $ & $ 0.8 \pm 0.2 $ & $ 1.4 \pm 0.4 $ & $ 2.1 \pm 0.5 $ & $ - $ \\

 \hline
 \end{tabular}
 \label{tab:main}
 \end{table}
 
\end{landscape}

	\subsection{Source properties}

We apply the source finding template to our five MWA images to perform aperture photometry and extract the peak, integrated, and background flux density and spectral index. We characterise the source centre and size of the major and minor axes by computing the spatial moments (or ``inertial axes'') of each source in the 216\,MHz image. We determine the position angle of the major axes, which allows us to define the HII regions as an ellipse. The integrated source and background flux density is derived by extracting the flux density within the template regions. We then subtract the corresponding background level to arrive at the background subtracted integrated flux density of the HII region. We determine the spectral index for each source by fitting the background subtracted integrated flux density across our five MWA bands using a least-squares fit to the function $S_{\rm \nu}=a\nu^{\alpha}$. Using this criteria we are able to determine the spectral index towards 184 HII regions and find an average spectral index of 1.5 with a standard error on the mean of 0.05. This is slightly lower than the expected values of 2.0, which suggests that we may have underestimated the contribution by the synchrotron background or that some HII regions are in the regime where the optical depth is beginning to turnover from optically thick to optically thin. We find 17 HII regions with spectral indices that are greater than 2.0 after including the uncertainty. Most of these HII regions (12) are absorption features in the 88\,MHz band, which would account for the steeper than expected spectral index. The high spectral index for the remaining five HII regions is mostly likely due to an overestimation of the background contribution in the lower frequency bands. For comparison the average spectral index of the raw integrated flux density estimates for our HII regions is much shallower at $\alpha = 0.4$. We also determine the spectral index for the background estimate and find an average value of $-0.7$ as expected for optically thin synchrotron emission.

\begin{figure}
\centering
\includegraphics[width=0.45\textwidth]{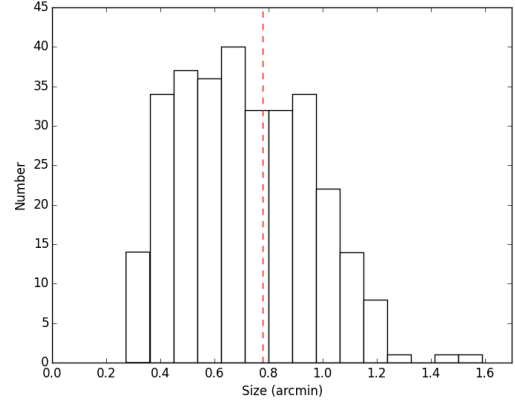} 
\caption{The angular size distribution for the sample of HII regions identified in our MWA three-colour images. The mean angular size of 6.0\arcmin\ is shown by a dashed red line.}
\label{im:sizes}
\end{figure}

We define the angular size of a source using the geometric mean of the ellipsoid major and minor axis given in columns 4 and 5 of Table~\ref{tab:main}. The angular size of the MWA HII regions ranges from 2 to 39\arcmin\ with a mean of 6.0\arcmin. The distribution of angular sizes can be seen in Fig.~\ref{im:sizes}. 

\section{Discussion}
\label{sec:disc}

	\subsection{HII regions blended with SNRs}
We find four cases where the colours in our MWA image suggest that HII regions are blended with SNRs that have not been previously identified. The HII regions associated with these sources are G333.6-0.02, G326.98+0.01, G320.37-1.05, and G318.28+0.09. These HII regions may be projected in front or behind the SNR emission or may be physically related. We find that these blended HII regions have lower than average spectral indices: 0.75, 0.98, 0.14, 0.48, for G333.6-0.02, G326.98+0.01, G320.37-1.05, and G318.28+0.09, respectively. This suggests that the integrated flux densities are being contaminated by the synchrotron emission from the SNRs. The SNRs associated with HII regions G333.6-0.02 and G326.98+0.01 are new detections and are presented along with other SNRs detected in these MWA images and associated properties in our partner publication \cite{Hollitt2015}.

	\subsection{The turnover frequency}\label{subsect:thick}

We expect to detect HII regions from the compact to classical stage. Such HII regions have typical sizes of $s\approx0.5$--10\,pc, electron temperatures of $T_{\rm e}\approx10^4$\,K and electron densities of $n_{\rm e}\approx 10^2$--$10^3$\,cm$^{-3}$ \citep{Kurtz2005}. The SED of HII regions with these properties turns over from optically thin to thick above 200\,MHz. If the turnover frequency ($\nu_{\rm t}$) can be identified it can be used to probe the physical properties of the HII region. From \cite{Mezger1967} we can write the turnover frequency of an HII region as:
\begin{equation}
\nu_{\rm t} = \left [ 0.082\,T_{\rm e}^{-1.35}\int n_{\rm e}^2\,ds \right ]^{0.476}
\end{equation}

\noindent Using the turnover frequency and the size of an HII region and assuming spherical geometry we can estimate the emission measure (EM: $\int n_{\rm e}^2\,ds$), and if the distance is known we may then determine the electron density, and thereby the ionised gas mass and ionising (Lyman continuum) photon count (N$_{\rm ly}$). 

We do not see any convincing evidence of a turnover for HII regions in our 88 to 216\,MHz band. To search for evidence of turnover at higher frequency we obtain the integrated flux density at 843 and 1420\,MHz using the MGPS and SGPS, respectively. The MGPS and SGPS surveys both suffer from significant artefacts and a lack of large scale sensitivity to structures greater $\sim30$\arcmin. This leads to lower than expected integrated flux densities, which results in an optically thin spectral index that is steeper than expected and a turnover that will be at a lower frequency. We therefore consider the results of the following analysis to be lower limits. We extract the integrated flux density in the MGPS and SGPS by applying our MWA source finding template. The resultant integrated fluxes can be seen in online version of Table~\ref{tab:main}. A number of our HII regions extend beyond the boundaries, or in some cases are outside of the SGPS and MGPS survey regions. We are unable to obtain an accurate estimate of the integrated flux density for these sources and so exclude them from further analysis. We find a total of 234 HII regions with complete MGPS and SGPS coverage. Where distances are available from the catalogue of HII regions in \cite{Anderson2014} we are able to use the turnover frequency to extract the physical properties mentioned above. We use the {\sc skyellipse} cross matching available in {\sc stilts} to determine the best matches between our MWA HII regions and those in \cite{Anderson2014} and check the results by eye. We find reliable distances to 61 HII regions which range from 1.6 to 14.7\,kpc. Using these distances we derive the physical radius of our HII region sample. We determine the electron temperature by cross matching our HII regions with those in \cite{Caswell1987}. If the electron temperature is not available we assume a representative electron temperature of $10^4$\,K \citep{Spitzer1950,Caswell1987}. The physical parameters derived using the turnover frequency are weakly dependant on the electron temperature and this assumption leads to at most a 10\% uncertainty in the derived physical properties. Under these limitations we are able to determine the turnover and physical properties towards 61 HII regions presented in Table~\ref{tab:to}. An important caveat is that due to the low resolution of these observations we will in many cases blend distinct HII regions into a single source. In the case of HII regions that are physically associated we will therefore estimate the average physical properties over the region. Sources that are located along the same line of site but are not physically associated will result in a contaminated estimate of the physical properties.  

\begin{figure}
\centering
\includegraphics[width=0.5\textwidth]{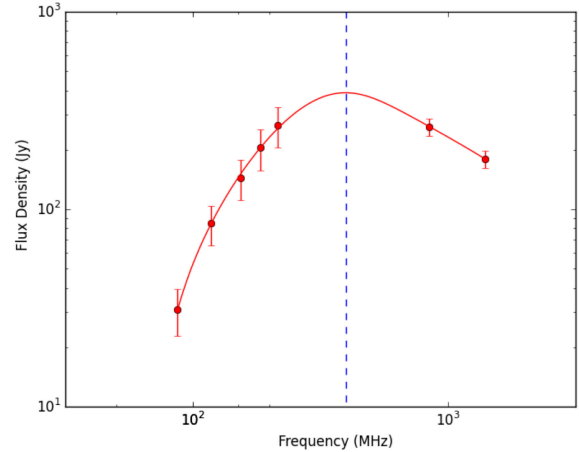} 
\caption{We show the turnover frequency of the HII region G305. The integrated flux density is shown by red symbols. The red line shows the best fit to the data. The vertical blue dashed line shows the turnover frequency of $390\pm30$\,MHz.}
\label{im:to}
\end{figure}

As an example we derive the turnover frequency for the well known G305 complex \cite{Hindson2013}. We find the total integrated flux density of the G305 complex by adding together the integrated flux density of the component HII regions. We perform a least-square fit to the SED between 88 and 1420\,MHz and find spectral indices of $2.3\pm0.3$ and $-0.5\pm0.2$ for the optically thick and thin regime, respectively (Fig.~\ref{im:to}). As expected the optically thin regime is steeper than the expected value of $-0.1$ due to the missing flux in the MGPS and SGPS surveys. For G305 we measure a turnover frequency of $\nu_{\rm t}=390\pm30$\,MHz. If we assume G305 has a spectral index of $-0.1$ in the optically thin regime this suggests that the MGPS and SGPS are missing approximately 20 and 60\% of the flux, respectively, at scales of $\approx1.0$\degree. The distance to G305 is $3.8\pm0.6$\,kpc \citep{Davies2012} giving a physical radius of $36\pm6$\,pc and the average electron temperature is 5400\,K. Using the electron temperature and turnover frequency we derive an emission measure of $1.8\pm0.3\times10^5$\,pc\,cm$^{-6}$. This leads to an electron density of $70.0\pm0.8$\,cm$^{-3}$, ionised gas mass of $3.5\pm1.7\times10^5$\,\msun\ and Lyman continuum photon rate of $1.0\pm0.5\times10^{50}$\,s$^{-1}$. This ionising flux is lower than that found in higher frequency observations at 5.5\,GHz ($2.4\times10^{50}$\,s$^{-1}$: \citealt{Hindson2013}) as is expected given the missing flux in the MGPS and SGPS surveys.

\begin{table*}
 \caption{Physical properties derived from our turnover frequency estimate. We caution that these results are lower limits due to the missing flux in the SGPS and MGPS.}
\scriptsize
\begin{tabular}{l|cccc|cccc}
\hline 
Name & $\nu_{\rm t}$ & Distance & T$_{\rm e}$ & Radius & EM & n$_{\rm e}$ & Mass & N$_{\rm ly}$\\
     & (MHz) & (kpc) & (K) & (pc) & (10$^5$pc\,cm$^{-6}$) & (10$^2$cm$^{-3}$) & (10$^3$\,\msun) & (10$^{48}$\,s$^{-1}$) \\
\hline
G284.73+0.32 & $ 730 \pm 65 $ & $ 6.0 \pm 1.3 $ & $ 6400 $ & $ 6.4 \pm 1.4 $ & $ 8.7 \pm 1.1 $ & $ 3.7 \pm 0.5 $ & $ 10.0 \pm 6.6 $ & $ 3.1 \pm 2.0 $\\
G$286.19-0.16$ & $ 280 \pm 43 $ & $ 2.4 \pm 0.1 $ & $ 7100 $ & $ 6.5 \pm 0.3 $ & $ 1.3 \pm 0.1 $ & $ 1.4 \pm 0.1 $ & $ 3.9 \pm 0.5 $ & $ 1.2 \pm 0.2 $\\
G286.43-1.48 & $ 650 \pm 98 $ & $ 9.2 \pm 1.4 $ & $ 8400 $ & $ 9.3 \pm 1.4 $ & $ 9.8 \pm 0.9 $ & $ 3.2 \pm 0.3 $ & $ 26.9 \pm 12.5 $ & $ 8.3 \pm 3.9 $\\
G287.25+0.36 & $ 600 \pm 53 $ & $ 2.5 \pm 0.1 $ & $ 6800 $ & $ 2.8 \pm 0.1 $ & $ 6.2 \pm 0.7 $ & $ 4.7 \pm 0.3 $ & $ 1.1 \pm 0.1 $ & $ 0.3 \pm 0.1 $\\
G289.08-0.36 & $ 590 \pm 68 $ & $ 7.9 \pm 1.2 $ & $ 8500 $ & $ 17.7 \pm 2.7 $ & $ 8.0 \pm 0.7 $ & $ 2.1 \pm 0.2 $ & $ 122.0 \pm 56.5 $ & $ 37.8 \pm 17.5 $\\
G289.77-1.16 & $ 340 \pm 34 $ & $ 8.4 \pm 1.2 $ & $ 9500 $ & $ 38.4 \pm 5.5 $ & $ 2.9 \pm 0.2 $ & $ 0.9 \pm 0.1 $ & $ 507.4 \pm 220.9 $ & $ 157.1 \pm 68.6 $\\
G291.30-0.70 & $ 920 \pm 78 $ & $ 3.1 \pm 0.1 $ & $ 7500 $ & $ 7.5 \pm 0.2 $ & $ 17.4 \pm 1.8 $ & $ 4.8 \pm 0.3 $ & $ 21.0 \pm 2.3 $ & $ 6.5 \pm 0.7 $\\
G291.60-0.50 & $ 720 \pm 190 $ & $ 7.7 \pm 1.2 $ & $ 6900 $ & $ 26.9 \pm 4.2 $ & $ 9.3 \pm 1.1 $ & $ 1.9 \pm 0.2 $ & $ 375.7 \pm 178.9 $ & $ 116.3 \pm 55.5 $\\
G291.90-0.97 & $ 290 \pm 30 $ & $ 8.2 \pm 1.2 $ & $ 6900 $ & $ 19.2 \pm 2.8 $ & $ 1.4 \pm 0.2 $ & $ 0.8 \pm 0.1 $ & $ 62.1 \pm 27.8 $ & $ 19.2 \pm 8.6 $\\
G297.53-0.77 & $ 430 \pm 30 $ & $ 10.2 \pm 1.2 $ & $ 6100 $ & $ 25.8 \pm 3.0 $ & $ 2.6 \pm 0.3 $ & $ 1.0 \pm 0.1 $ & $ 179.5 \pm 65.1 $ & $ 55.6 \pm 20.2 $\\
G298.53-0.35 & $ 150 \pm 15 $ & $ 10.5 \pm 1.1 $ & $ 6400 $ & $ 34.6 \pm 3.6 $ & $ 0.3 \pm 0.1 $ & $ 0.3 \pm 0.1 $ & $ 124.5 \pm 40.3 $ & $ 38.5 \pm 12.5 $\\
G299.03+0.14 & $ 200 \pm 20 $ & $ 10.6 \pm 1.1 $ & $ 6100 $ & $ 29.8 \pm 3.1 $ & $ 0.5 \pm 0.1 $ & $ 0.4 \pm 0.1 $ & $ 115.9 \pm 37.2 $ & $ 35.9 \pm 11.6 $\\
G299.37-0.26 & $ 530 \pm 45 $ & $ 4.2 \pm 0.1 $ & $ 10600 $ & $ 3.4 \pm 0.1 $ & $ 8.6 \pm 0.6 $ & $ 5.0 \pm 0.2 $ & $ 2.0 \pm 0.2 $ & $ 0.6 \pm 0.1 $\\
G302.68+0.17 & $ 320 \pm 30 $ & $ 4.6 \pm 1.3 $ & $ 5000 $ & $ 18.2 \pm 5.2 $ & $ 1.1 \pm 0.2 $ & $ 0.8 \pm 0.1 $ & $ 48.5 \pm 41.8 $ & $ 15.0 \pm 13.0 $\\
G305.21-0.37 & $ 200 \pm 20 $ & $ 3.8 \pm 0.6 $ & $ 4400 $ & $ 10.1 \pm 1.6 $ & $ 0.3 \pm 0.1 $ & $ 0.6 \pm 0.1 $ & $ 6.2 \pm 3.0 $ & $ 1.9 \pm 0.9 $\\
G305.23+0.03 & $ 470 \pm 28 $ & $ 3.8 \pm 0.6 $ & $ 5400 $ & $ 9.1 \pm 1.4 $ & $ 2.7 \pm 0.4 $ & $ 1.7 \pm 0.2 $ & $ 13.5 \pm 6.5 $ & $ 4.2 \pm 2.0 $\\
G305.25+0.22 & $ 540 \pm 33 $ & $ 3.8 \pm 0.6 $ & $ 5900 $ & $ 7.0 \pm 1.1 $ & $ 4.1 \pm 0.5 $ & $ 2.4 \pm 0.3 $ & $ 8.6 \pm 4.2 $ & $ 2.7 \pm 1.3 $\\
G305.39+0.20 & $ 520 \pm 45 $ & $ 3.8 \pm 0.6 $ & $ 5100 $ & $ 6.2 \pm 1.0 $ & $ 3.1 \pm 0.5 $ & $ 2.3 \pm 0.2 $ & $ 5.5 \pm 2.7 $ & $ 1.7 \pm 0.8 $\\
G305.56+0.36 & $ 290 \pm 30 $ & $ 3.8 \pm 0.6 $ & $ 5100 $ & $ 4.1 \pm 0.6 $ & $ 0.9 \pm 0.1 $ & $ 1.5 \pm 0.2 $ & $ 1.1 \pm 0.5 $ & $ 0.3 \pm 0.2 $\\
G305.70+1.63 & $ 650 \pm 98 $ & $ 3.8 \pm 0.6 $ & $ 6700 $ & $ 6.9 \pm 1.1 $ & $ 7.2 \pm 0.8 $ & $ 3.2 \pm 0.3 $ & $ 11.1 \pm 5.3 $ & $ 3.4 \pm 1.7 $\\
G305.82+0.16 & $ 310 \pm 23 $ & $ 3.8 \pm 0.6 $ & $ 7800 $ & $ 4.3 \pm 0.7 $ & $ 1.9 \pm 0.2 $ & $ 2.1 \pm 0.2 $ & $ 1.7 \pm 0.8 $ & $ 0.5 \pm 0.3 $\\
G306.35-0.36 & $ 710 \pm 58 $ & $ 8.9 \pm 0.8 $ & $ 4600 $ & $ 7.2 \pm 0.6 $ & $ 5.2 \pm 0.9 $ & $ 2.7 \pm 0.3 $ & $ 10.1 \pm 2.9 $ & $ 3.1 \pm 0.9 $\\
G309.10+0.17 & $ 370 \pm 33 $ & $ 5.4 \pm 1.2 $ & $ 4400 $ & $ 6.9 \pm 1.5 $ & $ 1.2 \pm 0.2 $ & $ 1.3 \pm 0.2 $ & $ 4.6 \pm 3.1 $ & $ 1.4 \pm 1.0 $\\
G309.57-0.72 & $ 710 \pm 110 $ & $ 5.4 \pm 2.5 $ & $ 5900 $ & $ 4.8 \pm 2.2 $ & $ 7.2 \pm 1.0 $ & $ 3.9 \pm 0.9 $ & $ 4.4 \pm 6.2 $ & $ 1.4 \pm 1.9 $\\
G309.92+0.39 & $ 460 \pm 40 $ & $ 5.4 \pm 0.1 $ & $ 6400 $ & $ 4.8 \pm 0.1 $ & $ 3.2 \pm 0.4 $ & $ 2.6 \pm 0.2 $ & $ 2.9 \pm 0.2 $ & $ 0.9 \pm 0.1 $\\
G310.20-0.12 & $ 510 \pm 33 $ & $ 11.5 \pm 0.9 $ & $ 10900 $ & $ 8.3 \pm 0.7 $ & $ 8.4 \pm 0.6 $ & $ 3.2 \pm 0.2 $ & $ 18.9 \pm 4.5 $ & $ 5.8 \pm 1.4 $\\
G310.99+0.42 & $ 410 \pm 20 $ & $ 5.6 \pm 1.2 $ & $ 5200 $ & $ 5.6 \pm 1.2 $ & $ 2.0 \pm 0.3 $ & $ 1.9 \pm 0.2 $ & $ 3.4 \pm 2.2 $ & $ 1.1 \pm 0.7 $\\
G311.14-0.26 & $ 470 \pm 50 $ & $ 14.7 \pm 1.2 $ & $ 9500 $ & $ 17.6 \pm 1.4 $ & $ 5.7 \pm 0.5 $ & $ 1.8 \pm 0.1 $ & $ 101.3 \pm 25.4 $ & $ 31.4 \pm 7.9 $\\
G311.23+0.78 & $ 390 \pm 33 $ & $ 5.6 \pm 0.1 $ & $ 7300 $ & $ 9.1 \pm 0.2 $ & $ 2.7 \pm 0.3 $ & $ 1.7 \pm 0.1 $ & $ 13.4 \pm 1.0 $ & $ 4.1 \pm 0.3 $\\
G314.26+0.43 & $ 610 \pm 45 $ & $ 5.9 \pm 0.5 $ & $ 5400 $ & $ 7.7 \pm 0.6 $ & $ 4.7 \pm 0.7 $ & $ 2.5 \pm 0.2 $ & $ 11.5 \pm 3.1 $ & $ 3.6 \pm 1.0 $\\
G316.39-0.37 & $ 200 \pm 23 $ & $ 12.7 \pm 0.8 $ & $ 5200 $ & $ 20.2 \pm 1.3 $ & $ 0.4 \pm 0.1 $ & $ 0.5 \pm 0.2 $ & $ 38.6 \pm 7.9 $ & $ 12.0 \pm 2.5 $\\
G316.81-0.05 & $ 680 \pm 50 $ & $ 2.5 \pm 0.6 $ & $ 5400 $ & $ 5.1 \pm 1.2 $ & $ 5.8 \pm 0.8 $ & $ 3.4 \pm 0.5 $ & $ 4.6 \pm 3.3 $ & $ 1.4 \pm 1.0 $\\
G319.91+0.78 & $ 550 \pm 30 $ & $ 2.6 \pm 0.5 $ & $ 5700 $ & $ 1.9 \pm 0.4 $ & $ 4.1 \pm 0.6 $ & $ 4.6 \pm 0.5 $ & $ 0.3 \pm 0.2 $ & $ 0.1 \pm 0.1 $\\
G320.19+0.80 & $ 520 \pm 27 $ & $ 2.4 \pm 0.5 $ & $ 9000 $ & $ 2.0 \pm 0.4 $ & $ 6.7 \pm 0.6 $ & $ 5.7 \pm 0.6 $ & $ 0.5 \pm 0.3 $ & $ 0.2 \pm 0.1 $\\
G320.28+0.43 & $ 410 \pm 23 $ & $ 2.1 \pm 0.6 $ & $ 11600 $ & $ 1.9 \pm 0.6 $ & $ 5.6 \pm 0.4 $ & $ 5.4 \pm 0.8 $ & $ 0.4 \pm 0.4 $ & $ 0.1 \pm 0.1 $\\
G320.72+0.23 & $ 450 \pm 140 $ & $ 12.8 \pm 0.7 $ & $ 7200 $ & $ 24.4 \pm 1.3 $ & $ 3.6 \pm 0.4 $ & $ 1.2 \pm 0.1 $ & $ 181.5 \pm 31.7 $ & $ 56.2 \pm 9.8 $\\
G321.06-0.50 & $ 400 \pm 20 $ & $ 4.1 \pm 0.6 $ & $ 5100 $ & $ 4.3 \pm 0.6 $ & $ 1.8 \pm 0.3 $ & $ 2.0 \pm 0.2 $ & $ 1.7 \pm 0.8 $ & $ 0.5 \pm 0.2 $\\
G321.15-0.55 & $ 1200 \pm 62 $ & $ 3.8 \pm 0.5 $ & $ 4500 $ & $ 4.8 \pm 0.6 $ & $ 14.8 \pm 2.6 $ & $ 5.6 \pm 0.6 $ & $ 6.4 \pm 2.6 $ & $ 2.0 \pm 0.8 $\\
G324.19+0.24 & $ 610 \pm 33 $ & $ 6.9 \pm 0.1 $ & $ 6800 $ & $ 8.6 \pm 0.1 $ & $ 6.3 \pm 0.7 $ & $ 2.7 \pm 0.2 $ & $ 17.9 \pm 1.3 $ & $ 5.6 \pm 0.4 $\\
G326.23+0.72 & $ 770 \pm 45 $ & $ 3.0 \pm 0.4 $ & $ 5000 $ & $ 10.9 \pm 1.5 $ & $ 6.9 \pm 1.1 $ & $ 2.5 \pm 0.3 $ & $ 33.8 \pm 13.9 $ & $ 10.5 \pm 4.3 $\\
G327.79-0.36 & $ 820 \pm 120 $ & $ 4.6 \pm 0.5 $ & $ 4800 $ & $ 5.2 \pm 0.6 $ & $ 7.5 \pm 1.2 $ & $ 3.8 \pm 0.4 $ & $ 5.5 \pm 1.9 $ & $ 1.7 \pm 0.6 $\\
G327.83+0.09 & $ 140 \pm 15 $ & $ 7.2 \pm 0.4 $ & $ 7500 $ & $ 14.8 \pm 0.8 $ & $ 0.3 \pm 0.1 $ & $ 0.5 \pm 0.2 $ & $ 15.5 \pm 2.7 $ & $ 4.8 \pm 0.8 $\\
G329.36+0.12 & $ 410 \pm 17 $ & $ 7.3 \pm 0.1 $ & $ 7300 $ & $ 7.7 \pm 0.1 $ & $ 3.1 \pm 0.3 $ & $ 2.0 \pm 0.1 $ & $ 9.3 \pm 0.6 $ & $ 2.9 \pm 0.2 $\\
G330.70-0.40 & $ 450 \pm 17 $ & $ 4.0 \pm 0.4 $ & $ 5000 $ & $ 4.5 \pm 0.4 $ & $ 2.2 \pm 0.3 $ & $ 2.2 \pm 0.2 $ & $ 2.0 \pm 0.6 $ & $ 0.6 \pm 0.2 $\\
G330.89-0.37 & $ 770 \pm 42 $ & $ 3.7 \pm 0.4 $ & $ 4900 $ & $ 11.4 \pm 1.2 $ & $ 6.7 \pm 1.1 $ & $ 2.4 \pm 0.2 $ & $ 37.6 \pm 12.7 $ & $ 11.6 \pm 3.9 $\\
G331.15-0.52 & $ 340 \pm 13 $ & $ 4.3 \pm 0.4 $ & $ 4500 $ & $ 8.9 \pm 0.8 $ & $ 1.1 \pm 0.2 $ & $ 1.1 \pm 0.1 $ & $ 7.9 \pm 2.3 $ & $ 2.5 \pm 0.7 $\\
G332.18-0.45 & $ 610 \pm 380 $ & $ 3.7 \pm 0.4 $ & $ 5600 $ & $ 3.7 \pm 0.4 $ & $ 4.9 \pm 0.7 $ & $ 3.7 \pm 0.3 $ & $ 1.9 \pm 0.6 $ & $ 0.6 \pm 0.2 $\\
G332.69-0.63 & $ 410 \pm 38 $ & $ 3.3 \pm 0.4 $ & $ 5200 $ & $ 6.3 \pm 0.8 $ & $ 1.9 \pm 0.3 $ & $ 1.7 \pm 0.2 $ & $ 4.5 \pm 1.7 $ & $ 1.4 \pm 0.5 $\\
G332.80-0.59 & $ 1100 \pm 58 $ & $ 3.8 \pm 0.4 $ & $ 5100 $ & $ 8.4 \pm 0.9 $ & $ 15.0 \pm 2.3 $ & $ 4.2 \pm 0.4 $ & $ 26.2 \pm 8.6 $ & $ 8.1 \pm 2.7 $\\
G333.04+2.03 & $ 820 \pm 98 $ & $ 1.6 \pm 0.6 $ & $ 6100 $ & $ 2.9 \pm 1.1 $ & $ 10.3 \pm 1.3 $ & $ 6.0 \pm 1.2 $ & $ 1.5 \pm 1.7 $ & $ 0.5 \pm 0.5 $\\
G333.64-0.22 & $ 480 \pm 18 $ & $ 3.2 \pm 0.4 $ & $ 6200 $ & $ 6.9 \pm 0.9 $ & $ 3.4 \pm 0.4 $ & $ 2.2 \pm 0.2 $ & $ 7.4 \pm 2.8 $ & $ 2.3 \pm 0.9 $\\
G333.71-0.46 & $ 370 \pm 10 $ & $ 11.8 \pm 0.4 $ & $ 2500 $ & $ 12.8 \pm 0.4 $ & $ 0.6 \pm 0.2 $ & $ 0.7 \pm 0.1 $ & $ 14.6 \pm 2.7 $ & $ 4.5 \pm 0.8 $\\
G336.56-0.20 & $ 250 \pm 25 $ & $ 10.4 \pm 0.4 $ & $ 7000 $ & $ 34.5 \pm 1.3 $ & $ 1.0 \pm 0.1 $ & $ 0.5 \pm 0.2 $ & $ 233.0 \pm 30.1 $ & $ 72.2 \pm 9.4 $\\
G336.82+0.04 & $ 390 \pm 43 $ & $ 7.8 \pm 0.9 $ & $ 6200 $ & $ 23.1 \pm 2.7 $ & $ 2.2 \pm 0.3 $ & $ 1.0 \pm 0.1 $ & $ 124.9 \pm 44.4 $ & $ 38.7 \pm 13.8 $\\
G337.16-0.16 & $ 510 \pm 48 $ & $ 10.9 \pm 0.4 $ & $ 5300 $ & $ 25.1 \pm 0.9 $ & $ 3.1 \pm 0.5 $ & $ 1.1 \pm 0.1 $ & $ 181.7 \pm 24.3 $ & $ 56.3 \pm 7.5 $\\
G337.63-0.07 & $ 290 \pm 33 $ & $ 11.8 \pm 0.4 $ & $ 4900 $ & $ 21.0 \pm 0.7 $ & $ 0.9 \pm 0.1 $ & $ 0.6 \pm 0.1 $ & $ 61.8 \pm 8.0 $ & $ 19.1 \pm 2.5 $\\
G338.95-0.08 & $ 1100 \pm 59 $ & $ 3.2 \pm 0.4 $ & $ 6000 $ & $ 2.2 \pm 0.3 $ & $ 19.7 \pm 2.6 $ & $ 9.5 \pm 0.9 $ & $ 1.0 \pm 0.4 $ & $ 0.3 \pm 0.1 $\\
G339.18-0.42 & $ 520 \pm 15 $ & $ 3.0 \pm 0.4 $ & $ 5600 $ & $ 3.5 \pm 0.5 $ & $ 3.6 \pm 0.5 $ & $ 3.2 \pm 0.3 $ & $ 1.4 \pm 0.6 $ & $ 0.4 \pm 0.2 $\\
G339.33+0.15 & $ 540 \pm 60 $ & $ 11.2 \pm 0.4 $ & $ 5100 $ & $ 16.1 \pm 0.6 $ & $ 3.4 \pm 0.5 $ & $ 1.5 \pm 0.1 $ & $ 62.3 \pm 8.3 $ & $ 19.3 \pm 2.6 $\\
G339.88+0.28 & $ 1100 \pm 180 $ & $ 14.2 \pm 0.6 $ & $ 6800 $ & $ 11.7 \pm 0.5 $ & $ 20.8 \pm 2.4 $ & $ 4.2 \pm 0.3 $ & $ 69.7 \pm 9.8 $ & $ 21.6 \pm 3.0 $\\
G340.33-0.21 & $ 370 \pm 30 $ & $ 3.5 \pm 0.4 $ & $ 5800 $ & $ 14.1 \pm 1.6 $ & $ 1.8 \pm 0.2 $ & $ 1.1 \pm 0.1 $ & $ 32.6 \pm 11.5 $ & $ 10.1 \pm 3.6 $\\

 	\hline
 \end{tabular}
 \label{tab:to}
 \end{table*}

	\subsection{WISE Comparison}
Cross matching this initial catalogue of HII regions with all previous surveys such as those described in Section~\ref{sec:intro} is beyond the scope of this paper. We choose to compare our results to one of the most comprehensive catalogues of HII regions in the Galaxy to date, which has been compiled by \cite{Anderson2014} using WISE. The resolution (6--12\arcsec) and sensitivity ($\sim 0.2$\,mJy\,beam$^{-1}$) of WISE should in theory be able to detect all HII regions in the Galactic plane. The authors compiled their catalogue by searching for the characteristic MIR morphology of HII regions and then searched the literature for additional information to determine distances. The catalogue consists of more than 8000 sources, which are split into the following categories: known; group; candidate; radio quiet, and sources with no radio data available. The sample includes 1986 candidate HII regions where the MIR emission is coincident with radio continuum emission but no radio recombination line (RRL) or H$\alpha$ emission. There are also $4124$ radio quiet HII regions that are not associated with any radio continuum emission. Finally, there are 115 regions that lack any high quality radio data. By comparing our catalogue of HII regions detected with the MWA with the catalogue of \cite{Anderson2014} we are able to check the validity of our HII region selection approach. We are also able to comment on the completeness of our sample and search for radio emission associated with radio quiet and no radio data HII regions in the WISE sample. 

There are 2247 HII regions detected by \cite{Anderson2014} between Galactic longitudes 260 and 340\degree. These HII regions consist of 260 known, 166 group, 818 candidate, 933 radio quiet sources which have no detectable radio emission, and 70 sources which have no available data at radio frequencies.  We use ellipse cross matching provided by {\sc stilts} \citep{Taylor2006} to match HII regions in our sample and the WISE catalogue. We then verify these matched regions by eye to ensure that we do not falsely match sources such as cases where unresolved WISE sources are embedded within larger HII regions. We find that all of the HII regions in our sample are associated with HII regions reported in the WISE sample. We do not identify any new HII regions however, which given our sensitivity and resolution is not surprising. The fact that all of our HII regions are identified in the WISE sample suggests that our three-colour selection method is effective at identifying optically thick HII regions. 

We use the \cite{Anderson2014} catalogue to search for optically thin HII regions in our images that may be missed due to our selection criteria. Such optically thin HII regions would have approximately flat ($\alpha = -0.1$) spectral indices and so appear white in our three-colour image described in Section~\ref{subsect:ident}. We do not find any convincing examples where an HII region identified by \cite{Anderson2014} is coincident with a source in our three-colour image that is indicative of an optically thin HII region. This suggests that the all of HII regions detected by our MWA observations are optically thick as we expect given the sensitivity, resolution and frequency range of the MWA, which limits our sample to bright and large compact and classical HII regions. 

We are in a position to search for emission associated with WISE HII regions that are classified as radio quiet and sources which have previously had no radio data available. For the 933 radio quiet WISE HII regions we are only able to detect low frequency radio emission associated with one HII region in the WISE sample which is designated G282.842-01.252 and G282.84-1.25 in the WISE and the MWA catalogue, respectively. We find that only 36 of the sample of radio quiet HII regions are reported to have sizes equal to or greater than our synthesised beam at 216\,MHz. It is therefore unsurprising that we do not detect many radio quiet sources. For the 70 sources with no radio data we are able to reliably identify radio emission in seven cases. 

The major limitation of our study is clearly the low resolution of our observations. Only 503 of the 2247 HII regions detected by WISE in our FoV would be at or above the resolution threshold in our 216\,MHz image. A significant number of these HII regions would also be blended in complex regions. If we compare our sample of 302 HII regions to the resolved WISE sample we arrive at a lower limit to the completeness of 61\%. Clearly a large number of HII regions we detect resolved as multiple HII regions by WISE. 

The majority of massive star formation occurs within a small number of massive star forming complexes. Many of our HII regions are associated with Group sources from \cite{Anderson2014}. A catalog of 88 massive star forming complexes using the Wilkinson Microwave Anisotropy Probe (WMAP) has also been compiled by \cite{Murray2010}. They report that over half of the total ionising luminosity of the Galaxy is produced within just 17 massive star forming complexes. There are 20 such massive star forming complexes in our survey region. We find HII regions associated 16 of these regions.

\section{Summary and Future Work}
\label{sec:conc}

This paper presents an initial sample of 302 HII regions detected in the GLEAM survey between $340 <l < 260$\degree. We exploit the wide-area, low frequency sky coverage of the MWA to detect HII regions without the need for ancillary data, which greatly simplifies source characterisation. The MWA frequency coverage allows us to distinguish morphologically similar but physically distinct objects by probing the significantly different SEDs of sources in the Galactic plane.

The GLEAM survey has observed the Galactic plane from 72 to 231\,MHz, covering the Galactic plane from $70 < l < 180$\degree\ and is currently being processed. The GLEAM pipeline includes peeling of bright sources in the primary beam sidelobes, correction for $w$-projection and ionospheric offsets and phase-only self-calibration. This will improve the astrometry and image fidelity of GLEAM images but we do not expect many more HII regions within the area presented here to be detected. 

The primary limitation of current MWA observations of Galactic HII regions is the low angular resolution. Current MWA observations are only able to resolve emission that is larger than approximately 2.5--5.6\arcmin\ depending on frequency. In the future both the upgraded MWA and further afield SKA1-LOW will provide an improvement in resolution that will allow us to detect many more HII regions in the Galactic plane. The expanded MWA will have approximately double the resolution of the current array and will allow us to improve our sensitivity. Such an improvement would allow us to resolve individual HII regions and identify both fainter and younger HII regions. The resolution of SKA1-LOW may even be sufficient to start to make such unambiguous detections in other, nearby galaxies. Additionally, the wider frequency coverage of SKA1-LOW (50--350\,MHz) will allow the turnover frequency of more HII regions to be detected.

\section{Acknowledgments}
LH was partially supported in this work via grant MED E1799 (PI: Johnston-Hollitt) provided by the Ministry of Business, Employment \& Innovation, New Zealand. MJ-H acknowledges support from the Marsden Fund. This scientific work makes use of the Murchison Radio-astronomy Observatory, operated by CSIRO. We acknowledge the Wajarri Yamatji people as the traditional owners of the Observatory site. We acknowledge the iVEC Petabyte Data Store, the Initiative in Innovative Computing and the CUDA Center for Excellence sponsored by NVIDIA at Harvard University. The authors thank the referee for their very useful comments that resulted in the improvement of this paper.

\bibliographystyle{apj}
\bibliography{paper.bib}

\end{document}